\documentclass[aps,prb,amsmath,amssymb,twocolumn,superscriptaddress]{revtex4-2}

\usepackage{graphicx}
\usepackage{dcolumn}
\usepackage{bm}

\setcitestyle{numbers,square}

\usepackage{graphicx}% Include figure files
\usepackage{dcolumn}% Align table columns on decimal point
\usepackage{bm}% bold math
\usepackage{hyperref}% add hypertext capabilities

\begin{document}

%\title{Photoinduced Excitonic Order Enhancement in spin crossover strongly correlated electron systems}
%\title{Photoinduced Excitonic Order Enhancement and Magnetic Transitions in Spin Crossover Strongly Correlated Electron Systems}
\title{Photoinduced Magnetic Transitions and Excitonic Order Enhancement in Spin Crossover Strongly Correlated Electron Systems}

\author{Yu. S. Orlov}
\affiliation{Institute of Engineering Physics and Radio Electronics, Siberian Federal University, 660041 Krasnoyarsk, Russia}
\affiliation{Kirensky Institute of Physics, Federal Research Center KSC SB RAS, 660036 Krasnoyarsk, Russia}

\author{S. V. Nikolaev}
\affiliation{Institute of Engineering Physics and Radio Electronics, Siberian Federal University, 660041 Krasnoyarsk, Russia}
\affiliation{Kirensky Institute of Physics, Federal Research Center KSC SB RAS, 660036 Krasnoyarsk, Russia}

\date{\today}

\begin{abstract}
The effects associated with exciton Bose condensate formation in strongly correlated spin crossover systems are considered within the effective Hamiltonian obtained from the two-orbital Hubbard-Kanamori model. The collective excitations spectrum at various points of the temperature-crystal field phase diagram is calculated. The role of the electron-phonon interaction is discussed. The exciton and magnetic order photoenhancement (induction) in strongly correlated spin crossover systems a new mechanism based on the cooperative effect of electron-phonon and interatomic exchange interactions and the appearance of a massive collective phase mode is demonstrated.
\end{abstract}

\maketitle

\section{Introduction}

Spin crossover is a special type of switching from one magnetic state to another, or from magnetic to non-magnetic. Instead of changing the magnetization orientation used in the traditional magnetic memory approach, the spin crossover makes it possible to change the magnetization magnitude, and thereby realize the possibility of an alternative form of magnetic recording and reading of information. Of particular interest are the spin crossover possibilities under the action of femtosecond laser pulses in connection with the recent works on an all-optical scheme for magnetic recording and reading of information. In this work, we will consider a class of phenomena caused by optical pumping of multi-electron states in spin crossover strongly correlated systems and their photothermal instability. The paper consists of two parts. In the first part, in the framework of the effective Hamiltonian obtained from the two-band Hubbard-Kanamori model we will consider the features formation of an exciton condensate, which is the condensation of local (at a crystal lattice site) magnetic excitons, in strongly correlated systems near the spin crossover. The formation of such a condensate inevitably determines the magnetic properties. With the help of the generalized spin waves method \cite{Nasu2016}, the collective excitations spectrum at different points of the temperature-crystal field phase diagram is calculated without and taking into account the electron-phonon interaction. In the second part, excitonic order photoenhancement and magnetic transitions photoinduction is demonstrated. The study of strongly correlated systems nonequilibrium dynamics can provide new knowledge in understanding their properties and new ways to control various ordered states that arise as a result of phase transitions. For example, the nonequilibrium properties of superconductors are now widely studied. Both the light-induced superconductivity enhancement \cite{Science.331.2011, PhysRevB.89.184516, Nature.530.2016} and the Higgs mode observation \cite{PhysRevLett.111.057002, Science.345.2014, PhysRevB.96.020505} have been reported. Recently, a related family of ordered states, exciton dielectrics, has also attracted interest \cite{Nature.471.2011, NatCommun.3.1069, NatureMater.13.857, PhysRevB.94.035121, PhysRevLett.119.086401, Werdehausen_2018}. Until now, theoretical and experimental works on exciton dielectrics and exciton ordering in spin crossover systems \cite{Nasu2016, JPhysCondensMatter.27.333201, PhysRevLett.99.126405, PhysRevB.80.054410, PhysRevB.62.2346, PhysRevB.90.245144, PhysRevB.89.115134} have been focused on the study of their equilibrium properties. The study of nonequilibrium properties has just begun \cite{PhysRevB.94.035121}. Of all the works known to us in this direction, we would like to highlight the work \cite{PhysRevLett.119.247601}, which is quite close in content to the present work. The authors of \cite{PhysRevLett.119.247601} succeeded in demonstrating the photoenhancement of the exciton order in the framework of the two-band model of spinless fermions associated with phonons. According to the authors of \cite{PhysRevLett.119.247601}, the combination of photoexcitation and Hamiltonian symmetry reduction can provide a new strategy for increasing the superconducting order parameter in superconductors. Some similarity between the exciton condensate state and the superconducting one makes it interesting to study the nonequilibrium properties of such systems. The work's scientific novelty lies in the identification of cooperative effects caused by kinematic and electron-phonon interactions in non-equilibrium conditions.

The paper is organized as follows. In Sec.~\ref{sec:2} we describe the two-orbital Hubbard-Kanamori model, the effective low energy Hamiltonian containing HS and LS states, and interatomic exchange interaction and exciton hopping. The phase diagrams and collective excitations spectra for different parameter sets are discussed in Sec.~\ref{sec:3}. Equation of motion for density matrix solution results are presented in Sec.~\ref{sec:4}, describing the changes in the exciton condensate and magnetization under the action of external electromagnetic radiation. In Sec.~\ref{sec:5} we discuss the main results.

\section{\label{sec:2}Two-orbital Hubbard-Kanamori model and effective Hamiltonian}

A minimal model of strongly correlated spin crossover systems is the two-orbital Hubbard-Kanamori model. Its Hamiltonian can be written as
\begin {equation}
\hat H_{(2)} = \hat H_\Delta + \hat H_t  + \hat H_{Coulomb}.
\label {H_two_band}
\end {equation}
Here, the first term
\begin {equation}
\hat H_\Delta = \varepsilon _1 \sum\limits_{i,\gamma } {c_{1i\gamma }^\dag  c_{1i\gamma }^{ } }  + \varepsilon _2 \sum\limits_{i,\gamma } {c_{2i\gamma }^\dag  c_{2i\gamma }^{ } }
\label {H_Delta}
\end {equation}
contains the one-ion energy of one-particle electron states with the energy levels $\varepsilon _1 $ and $\varepsilon _2  = \varepsilon _1  + \Delta $, where $\Delta$ is the crystal field energy (for convenience one can assume $\varepsilon _1 = 0$), $c_{\lambda i\gamma }^\dag$ creates a fermion at orbital $\lambda = 1,2$, site $i$, and with spin projection $\gamma = \pm 1/2$. The second term is
\begin {eqnarray}
\hat H_t &=& t_{11} \sum\limits_{\left\langle {i,j} \right\rangle ,\gamma } {c_{1i\gamma }^\dag  c_{1j\gamma }^{ } }  + t_{22} \sum\limits_{\left\langle {i,j} \right\rangle ,\gamma } {c_{2i\gamma }^\dag  c_{2j\gamma }^{ } } \nonumber \\
 &+& t_{12} \sum\limits_{\left\langle {i,j} \right\rangle ,\gamma } {\left( {c_{2i\gamma }^\dag  c_{1j\gamma }^{ }  + c_{1i\gamma }^\dag  c_{2j\gamma }^{ } } \right)},
\label {H_t}
\end {eqnarray}
where $t_{\lambda \lambda '} $ is the nearest neighbor hopping parameter. The third term
\begin{widetext}
\begin{eqnarray}
\hat H_{Coulomb} = U\sum\limits_{\lambda, i } {c_{\lambda i \uparrow }^\dag  c_{\lambda i \downarrow }^\dag  c_{\lambda i \uparrow }^{ } c_{\lambda i \downarrow }^{ } }  + V\sum\limits_{\lambda \ne \lambda', i  } {c_{\lambda i \uparrow }^\dag  c_{\lambda' i \downarrow }^\dag  c_{\lambda i \uparrow }^{ } c_{\lambda' i \downarrow }^{ } } + V\sum\limits_{\lambda  > \lambda ',i, \gamma } {c_{\lambda i \gamma }^\dag  c_{\lambda' i \gamma }^\dag  c_{\lambda i \gamma }^{ } c_{\lambda' i \gamma }^{ } } \nonumber \\
+ J_H \sum\limits_{\lambda  > \lambda ', i, \gamma } {c_{\lambda i \gamma }^\dag  c_{\lambda' i \gamma }^\dag  c_{\lambda' i \gamma }^{ } c_{\lambda i \gamma }^{ } } + J_H \sum\limits_{\lambda \ne \lambda', i} {c_{\lambda i \uparrow }^\dag  c_{\lambda' i \downarrow }^\dag  c_{\lambda' i \uparrow }^{ } c_{\lambda i \downarrow }^{ } }  + J'_H \sum\limits_{\lambda \ne \lambda', i} {c_{\lambda i \uparrow }^\dag  c_{\lambda i  \downarrow }^\dag  c_{\lambda' i \uparrow }^{ } c_{\lambda' i \downarrow }^{ } }
\label {H_Coulomb}
\end{eqnarray}
\end{widetext}
includes the one-site Coulomb interaction energy. Electron-electron interaction is considered in the Kanamori approximation \cite{Kanamori1963}. It preserves the multiplet character of the electron-electron interaction as much as possible. $U$ is the parameter that describes the direct repulsion between two electrons in the same orbital. $V$ describes the direct repulsion between two electrons in different orbitals. It is assumed to be equal between all different orbitals. The process where two electrons are interchanged is described by the exchange integral $J_H$. The process where two electrons residing in the same orbital scatter on each other and are transferred from one orbital into another is characterized by the integral ${J'_H}$.

An important feature of such a two-orbital model is a possibility of formation, at half-filling ($N_e=2$ is an average number of electrons on a crystal lattice site) and in the zero hopping approximation $t_{\lambda\lambda'}=0$, of various localized two-electron states with spin values $S=0,1$, which makes possible a spin crossover with varying $\Delta$. Within the region $\Delta  < \Delta _C  = \sqrt {\left( {U - V + J_H} \right)^2  + {J_H'}^2 } $ the ground state is the triplet high-spin (HS) state ($S=1$) $\left| \sigma  \right\rangle$ with the energy $E_{HS}$:
\begin{eqnarray}
\left| \sigma  \right\rangle  = \left\{ \begin{array}{l}
 c_{1 \uparrow }^\dag  c_{2 \uparrow }^\dag  \left| 0 \right\rangle ,{\rm{   }}\sigma  =  + 1 \\
 \frac{1}{{\sqrt 2 }}\left( {c_{1 \uparrow }^\dag  c_{2 \downarrow }^\dag  \left| 0 \right\rangle  + c_{1 \downarrow }^\dag  c_{2 \uparrow }^\dag  \left| 0 \right\rangle } \right),{\rm{   }}\sigma  = 0\\
 c_{1 \downarrow }^\dag  c_{2 \downarrow }^\dag  \left| 0 \right\rangle ,{\rm{   }}\sigma  =  - 1, \\
 \end{array} \right.{\rm{   }} \nonumber
\end {eqnarray}
while at $\Delta  > \Delta _C $ the ground state is the low-spin (LS) singlet ($S=0$) state $\left| s \right\rangle  = C_1 \left( \Delta  \right)c_{1 \uparrow }^\dag  c_{1 \downarrow }^\dag  \left| 0 \right\rangle  - C_2 \left( \Delta  \right)c_{2 \uparrow }^\dag  c_{2 \downarrow }^\dag  \left| 0 \right\rangle $ with the energy $E_{LS} $, where $C_1 \left( \Delta  \right) = \sqrt {1 - C_2^2 \left( \Delta  \right)} $ and $C_2 \left( \Delta  \right) = {x \mathord{\left/ {\vphantom {x 2}} \right. \kern-\nulldelimiterspace} 2}\left( {1 + x + \sqrt {1 + x} } \right)$ are the normalizing coefficients \cite{Orlov_PhysRevB.104.195103} ($x = {{{J'_H}^2 } \mathord{\left/
 {\vphantom {{{J'_H}^2 } {\Delta ^2 }}} \right.
 \kern-\nulldelimiterspace} {\Delta ^2 }}$).

To obtain an effective Hamiltonian it is convenient to use Hubbard $X$-operators $X^{p,q}  = \left| p \right\rangle \left\langle q \right|$ \cite{Hubbard1964} built on the eigenstates of the Hamiltonian $\hat H_\Delta + \hat H_{Coulomb}$:
\begin{equation}
\left( {\hat H_\Delta + \hat H_{Coulomb} } \right)\left| p \right\rangle  = E_p \left| p \right\rangle
\label {HDelta_HCoulomb}
\end{equation}
with the number of electrons taking values $N_e = 1,2,3$. Since the Hubbard operators form a linearly independent basis, any local operator can be represented as a linear combination of the $X$-operators, including the one-electron annihilation operator
\begin{equation}
c_{\lambda i \gamma }  = \sum\limits_{pq} {\left| p \right\rangle \left\langle {p\left| {c_{\lambda i \gamma } } \right|q} \right\rangle \left\langle q \right|}  = \sum\limits_{pq} {\chi _{\lambda \gamma } \left( {p, q} \right)X_i^{p, q} }.
\label {c_in_X}
\end{equation}
Since the number of eigenstates $\left| p \right\rangle $ and $\left| q \right\rangle $ is finite, the pairs $\left( {p,q} \right)$ can be numbered by an index $m$ (or $n$) for convenience \cite{Zaitsev1976}. The notation of the diagonal $X$-operators $X^{p,p}$ remains unchanged.

Using Eq.~\ref{c_in_X}, the anomalous averages $\left\langle {c_{2f\gamma }^\dag  c_{1f\gamma }^{ } } \right\rangle $ without and with a spin projection change, $\left\langle {c_{2f\bar \gamma }^\dag  c_{1f\gamma }^{ } } \right\rangle $ ($\bar \gamma  =  - \gamma $), can be written as
\begin{equation}
\left\langle {c_{2f\gamma }^\dag  c_{1f\gamma }^{ } } \right\rangle  \approx  - \gamma \sqrt 2 \left( {C_2 \left\langle {X_f^{s,0} } \right\rangle  + C_1 \left\langle {X_f^{0,s} } \right\rangle } \right),
\label {anomal_1}
\end{equation}
\begin{eqnarray}
&\left\langle {c_{2f\bar \gamma }^\dag  c_{1f\gamma } } \right\rangle&  \approx  - 2\gamma \left( {\gamma  + \frac{1}{2}} \right)\left( {C_2 \left\langle {X_f^{s, + 1} } \right\rangle  + C_1 \left\langle {X_f^{ + 1,s} } \right\rangle } \right) \nonumber \\
 &+& 2\gamma \left( {\gamma  - \frac{1}{2}} \right)\left( {C_2 \left\langle {X_f^{s, - 1} } \right\rangle  + C_1 \left\langle {X_f^{ - 1,s} } \right\rangle } \right).
\label {anomal_2}
\end{eqnarray}
Here and below, angular brackets denote thermodynamical averages.

As follows from Eqs.~\ref{anomal_1},~\ref{anomal_2}, the exciton pairing is described by non-zero averages of singlet-triplet excitations. The Hamiltonian defined by Eq.~\ref{H_two_band} can be rewritten in the $X$-operator representation as
\begin{equation}
\hat H_{\left( 2 \right)} = \sum\limits_{i,p} {E_p X_i^{p,p} }  + \sum\limits_{\left\langle {i,j} \right\rangle } {\sum\limits_{mn} {t^{mn} X_i^{\dag m} X_j^n } },
\label {H_two_band_in_X}
\end{equation}
where $E_p $ is the multielectron eigenstate energy and $t^{mn}  = \sum\limits_{\lambda ,\lambda ',\gamma } {t_{\lambda \lambda '} \chi _{\lambda \gamma }^* \left( m \right)\chi _{\lambda '\gamma } \left( n \right)} $ is the renormalized hopping integral.

Using the Hamiltonian in Eq.~\ref{H_two_band_in_X} as an initial one, we can obtain an effective Hamiltonian by excluding interband (between the lower and upper Hubbard subbands) hopping in a second order perturbation theory over interatomic electron hopping similar to the t-J model derivation from the Hubbard model. To do this, we apply the projection operator method developed in Ref.~\cite{Chao1977} for the Hubbard model and in Ref.~\cite{Gavrichkov_Polukeev} for the $p-d$ model (see also Refs.~\cite{JPhysCondensMatter.27.333201, Nasu2016}).

The obtained effective Hamiltonian is
\begin{equation}
\hat H_{eff} = \hat H_S + \hat H_{n_{LS}n_{LS}} + \hat H_{ex}.
\label {H_eff}
\end{equation}
Here, the first term describes an AFM exchange contribution to the Heisenberg-like Hamiltonian
\begin{equation}
\hat H_S = \frac{1}{2}J\sum\limits_{\left\langle {i,j} \right\rangle } {\left( { \mbox{\boldmath{$\hat S$}}_i \cdot \mbox{\boldmath{$\hat S$}}_j  - \frac{1}{4} \hat n_i \hat n_j } \right)},
\label {H_S}
\end{equation}
where $\mbox{\boldmath{$\hat S$}}_i$ is the $S=1$ spin operator: $\hat S_i^ +   = \sqrt 2 \left( {X_i^{ + 1,0}  + X_i^{0, - 1} } \right)$, $\hat S_i^ -   = \sqrt 2 \left( {X_i^{0, + 1}  + X_i^{ - 1,0} } \right)$, and $\hat S_i^z  = X_i^{ + 1, + 1}  + X_i^{ - 1, - 1} $ \cite{Valkov}; $J = {{\left( {t_{11}^2  + 2t_{12}^2  + t_{22}^2 } \right)} \mathord{\left/
 {\vphantom {{\left( {t_{11}^2  + 2t_{12}^2  + t_{22}^2 } \right)} {\Omega _g }}} \right.
 \kern-\nulldelimiterspace} {\Omega _g }}$ is the interatomic exchange interaction, where $\Omega_g$ is the energy interval between the centers of the upper and lower Hubbard subbands (charge-transfer energy); $\hat n_i  = 2\left( {X_i^{s,s}  + \sum\limits_\sigma  {X_i^{\sigma ,\sigma } } } \right) = 2\left( {\hat n_i^{LS} + \hat n_i^{HS} } \right)$ is the particle number operator at site $i$ ($\hat n_i^{LS\left( {HS} \right)} $ is the occupation operator of the LS (HS) state). Using the completeness condition $X^{s,s}  + \sum\limits_\sigma  {X^{\sigma ,\sigma } }  = 1$, one can show that $\left\langle {\hat n_i } \right\rangle  = 2\left( {\left\langle {\hat n_i^{LS} } \right\rangle  + \left\langle {\hat n_i^{HS} } \right\rangle } \right) = 2\left( {n_{LS}  + n_{HS} } \right) = 2$.

The next term,
\begin{equation}
\hat H_{n_{LS}n_{LS}} = \frac{1}{2}\tilde J\sum\limits_{\left\langle {i,j} \right\rangle } {X_i^{s,s}  \cdot X_j^{s,s} },
\label {H_nn}
\end{equation}
where $\tilde J = \left[ {1 - \left( {2C_1 C_2 } \right)^2 } \right]{{\left( {t_{11}^2  - 2t_{12}^2  + t_{22}^2 } \right)} \mathord{\left/
 {\vphantom {{\left( {t_{11}^2  - 2t_{12}^2  + t_{22}^2 } \right)} {\Omega _g }}} \right.
 \kern-\nulldelimiterspace} {\Omega _g }}$, represents a density-density type interaction of LS states.

The third term in Eq.~\ref{H_eff},
\begin{widetext}
\begin{equation}
\hat H_{ex}  =  -\frac{{\varepsilon _S }}{2} \sum\limits_i \left({X_i^{s,s} }  - {\sum\limits_{\sigma  =  - S}^{ + S} {X_i^{\sigma ,\sigma } } } \right) + \sum\limits_\sigma  {\sum\limits_{\left\langle {i,j} \right\rangle } {\left[ {\frac{1}{2}J'_{ex} \left( {X_i^{\sigma ,s} X_j^{s,\sigma }  + X_i^{s,\sigma } X_j^{\sigma ,s} } \right) - \frac{1}{2}J''_{ex} ( - 1)^{\left| \sigma  \right|} \left( {X_i^{\sigma ,s} X_j^{\bar \sigma ,s}  + X_i^{s,\sigma } X_j^{s,\bar \sigma } } \right)} \right]} },
\label {H_ex}
\end{equation}
\end{widetext}
contains singlet and triplet energies as well as interatomic hoppings of excitons with the amplitude $J'_{ex}$ as well as creation/annihilation processes of biexcitons on neighboring sites with the amplitude $J''_{ex}$. In the absence of any cooperative interactions, at negative values of the spin gap $\varepsilon _S  = E_{HS}  - E_{LS} $ the ground state is the HS state, whereas at positive spin gap values, the ground states is the LS state; $J'_{ex}  = 2C_1 C_2 {{\left( {t_{11} t_{22}  - t_{12}^2 } \right)} \mathord{\left/
 {\vphantom {{\left( {t_{11} t_{22}  - t_{12}^2 } \right)} {\Omega _g }}} \right.
 \kern-\nulldelimiterspace} {\Omega _g }}$, $J''_{ex}  = {{\left( {t_{11} t_{22}  - t_{12}^2 } \right)} \mathord{\left/
 {\vphantom {{\left( {t_{11} t_{22}  - t_{12}^2 } \right)} {\Omega _g }}} \right.
 \kern-\nulldelimiterspace} {\Omega _g }}$, $\bar \sigma  =  - \sigma$. The Hubbard operators $X_i^{\sigma ,s}$ and $X_i^{s,\sigma }$ in Eq.~\ref{H_ex} describe Bose-like excitations (excitons) between the LS singlet state $\left| s \right\rangle$ and the HS triplet state $\left| \sigma  \right\rangle$. The first term within the square brackets in Eq.~\ref{H_ex} describes the excitonic dispersion by means of interatomic hoppings (such a dispersion was considered long ago in the work of Vonsovskii and Svirskii~\cite{Vonsovsky}). The second term in Eq.~\ref{H_ex} contains creation/annihilation processes of biexcitons at neighboring sites of a lattice, which makes the dispersion more complicated compared to the usual one obtained within the tight binding method~\cite{Vonsovsky}. Near the spin crossover, the normalization constants defined above take values $C_1  \approx 1$ and $C_2  \approx 0$, thus, $J'_{ex}  \approx 0$ \cite{Orlov_PhysRevB.104.195103}. At such circumstances, the biexciton excitations play the main role in the formation of the excitonic dispersion.

It is useful to introduce another set of the operators $\hat d_+  = X^{s, + }$, $\hat d_-  = X^{s, - }$, and $\hat d_0 = X^{s,0}$, and their combinations ${\hat d_x} = \frac{1}{{\sqrt 2 }}\left( { - {\hat d_ + } + {\hat d_ - }} \right)$, ${\hat d_y} = \frac{1}{{\sqrt 2 i}}\left( {{\hat d_ + } + {\hat d_ - }} \right)$, and ${\hat d_z} = {\hat d_0}$ \cite{PhysRevB.89.115134}. $\mbox{\boldmath{$\hat d$}} = \left( {{\hat d_x},{\hat d_y},{\hat d_z}} \right)$ corresponds to the "$\mbox{\boldmath{$d$}}$ vector" in the triplet superconductivity. The last term in Eq.~\ref{H_ex} can be performed in $\mbox{\boldmath{$\hat d$}}$ operators representation as
\begin{equation}
\frac{1}{2}{J'_{ex}}\sum\limits_{\left\langle {i,j} \right\rangle } {\left( {\mbox{\boldmath{$\hat d$}}_i^\dag  \cdot {\mbox{\boldmath{$\hat d$}}_j} + {\mbox{\boldmath{$\hat d$}}_i} \cdot \mbox{\boldmath{$\hat d$}}_j^\dag } \right)}  - \frac{1}{2}{J''_{ex}}\sum\limits_{\left\langle {i,j} \right\rangle } {\left( {\mbox{\boldmath{$\hat d$}}_i^\dag  \cdot \mbox{\boldmath{$\hat d$}}_j^\dag  + {\mbox{\boldmath{$\hat d$}}_i} \cdot {\mbox{\boldmath{$\hat d$}}_j}} \right)}.
\label {H_ex_in_d}
\end{equation}
The first term represents the exchange of the LS and HS states between the NN sites. The second term represents changes in the spin states in the NN sites, where the LS states are changed into the HS states in both the $i$ and $j$ sites and vice versa.

Taking into account the electron-phonon interaction one has
\begin{equation}
\hat H = \hat H_{eff} + \hat H_{1ph} + H_{2ph},
\label{H}
\end{equation}
where
\begin{eqnarray}
\hat H_{1ph} &=& {\omega _0} _{\left( 1 \right)} \sum\limits_i {\left( {a_i^\dag  a_i^{ }  + \frac{1}{2}} \right)} \nonumber \\
&+& {g_1}\sum\limits_i {\left( {{a_i} + a_i^\dag } \right)\left( {X_i^{s,s} - \sum\limits_{\sigma  =  - 1}^{ + 1} {X_i^{\sigma ,\sigma }} } \right)},
\label {H_1ph}
\end{eqnarray}
\begin{eqnarray}
\hat H_{2ph} &=& {\omega _0} _{\left( 2 \right)} \sum\limits_i {\left( {b_i^\dag  b_i^{ }  + \frac{1}{2}} \right)} \nonumber \\
&+& {g_2}\sum\limits_i {\sum\limits_{\sigma  =  - 1}^{ + 1} {\left( {{b_{i,\sigma }} + b_{i,\sigma }^\dag } \right)\left( {X_i^{s,\sigma } + X_i^{\sigma ,s}} \right)} }.
\label {H_2ph}
\end{eqnarray}
The term in Eq.~\ref{H_1ph} contains the diagonal electron-phonon interaction. Next, the term in Eq.~\ref{H_2ph} describes off-diagonal electron-phonon transition processes between the LS and HS states. Here, $g_{1\left( 2 \right)} $ are the constants of electron-phonon interaction, ${\omega _0} _{\left( {1,2} \right)} $ are the frequencies of ``a''- and ``b''-type phonons (${\omega _0} _{\left( {1,2} \right)} = 0.05 eV$ below).
%The terms proportional to $V_{a\left( b \right)} $ describe interactions of $a$($b$)-phonons at different sites of a crystal lattice.

\section{\label{sec:3}MF phase diagram and collective excitation spectra}

The Hamiltonian~\ref{H} is divided into the mean field (MF) term and the fluctuation term as $\hat H = {\hat H^{MF}} + \delta \hat H$.
In MF approximation for two sublattices $A$ and $B$, the terms in Eqs.~\ref{H_S}--\ref{H_ex} can be expressed as the following Eqs.~\ref{H_S_MF}--\ref{H_ex_MF}:

\begin{eqnarray}
\hat H_S^{MF} = zJm_B \sum\limits_{i_A } {\hat S_{i_A }^z }  + zJm_A \sum\limits_{i_B } {\hat S_{i_B }^z } \nonumber \\
 - zJ\frac{1}{4}n_B \sum\limits_{i_A } {\hat n_{i_A } }  - zJ\frac{1}{4}n_A \sum\limits_{i_B } {\hat n_{i_B } } \nonumber \\
  - \frac{1}{2}zJNm_A m_B  + \frac{1}{2}zJN,
\label {H_S_MF}
\end{eqnarray}
where $z$ is a number of nearest neighbors and $m_{A\left( B \right)}  = \left\langle {\hat S_{i_{A\left( B \right)} }^z } \right\rangle$ is an $A$($B$)-sublattice magnetization, $N$ is the number of crystal lattice sites;
\begin{eqnarray}
\hat H_{n_{LS}n_{LS}}^{MF} &=& z\tilde Jn_{LS,B} \sum\limits_{i_A } {\hat n_{i_A }^{LS} }  + z\tilde Jn_{LS,A} \sum\limits_{i_B } {\hat n_{i_B }^{LS} } \nonumber \\
&-& z\tilde J\frac{N}{2}n_{LS,A} n_{LS,B}.
\label {H_nn_MF}
\end{eqnarray}

The interaction proportional to $\tilde J$ leads to an additional cooperation mechanism, but in the following we will mainly neglect it to simplify the results, since it does not influence the behavior of phase diagrams qualitatively, leading only to the sublattices LS energies renormalization.

\begin{widetext}
\begin{eqnarray}
\hat H_{ex}^{MF}  = \sum\limits_F {\sum\limits_{\sigma  =  \pm 1,0} {\left\{ {zJ'_{ex} \Delta _{ex,\bar F}^\sigma  \sum\limits_{i_F } {\left( {X_{i_F }^{s,\sigma }  + X_{i_F }^{\sigma ,s} } \right)}  - \left( { - 1} \right)^{\left| \sigma  \right|} zJ''_{ex} \Delta _{ex,\bar F}^\sigma  \sum\limits_{i_F } {\left( {X_{i_F }^{s,\bar \sigma }  + X_{i_F }^{\bar \sigma ,s} } \right)} } \right.} } \nonumber  \\
\left. { - \frac{1}{2}zN\left( {J'_{ex} \Delta _{ex,F}^\sigma  \Delta _{ex,\bar F}^\sigma   - \left( { - 1} \right)^{\left| \sigma  \right|} J''_{ex} \Delta _{ex,F}^\sigma  \Delta _{ex,\bar F}^{\bar \sigma } } \right)} \right\} - \varepsilon _S \sum\limits_{i_A } {X_{i_A }^{s,s} }  - \varepsilon _S \sum\limits_{i_B } {X_{i_B }^{s,s} }  + N\frac{{\varepsilon _S }}{2},
\label {H_ex_MF}
\end{eqnarray}
\end{widetext}
where $F = \left(A,B\right)$ ($\bar F = A$ if $F = B$ and vice versa), $\Delta _{ex,A\left( B \right)}^\sigma   = \left\langle {X_{i_A \left( {i_B } \right)}^{s,\sigma } } \right\rangle $ are the excitonic order parameter components, which satisfy the equation $\left( {\Delta _{ex}^\sigma  } \right)^\dag   = \left\langle {X^{\sigma ,s} } \right\rangle  = \Delta _{ex}^\sigma $ at thermodynamic equilibrium. Note that, when $\Delta _{ex}^\sigma   \ne 0$, a quantum mechanical mixture of the LS and HS states is present, albeit in the absence of spin-orbital interaction.

By solving the eigenstate problem
\begin{equation}
{\hat H^{MF}}{\left| \psi_k  \right\rangle} = {E_k}{\left| \psi_k  \right\rangle},
\label {self_eig}
\end{equation}
where
\begin{widetext}
\begin{equation}
\left| {{\psi _k}} \right\rangle  = \sum\limits_{{n_{1ph}} = 0}^{{N_{1ph}}} {\sum\limits_{\sigma  =  - 1}^{ + 1} {\sum\limits_{{n_{2ph}} = 0}^{{N_{2ph}}} {\left[ {{C_{HS,k,\sigma }}\left( {{n_{1ph}},{n_{2ph}}} \right)\left| {\sigma ,{n_{1ph}},{n_{2ph}}} \right\rangle  + {C_{LS,k}}\left( {{n_{1ph}},{n_{2ph}}} \right)\left| {s,{n_{1ph}},{n_{2ph}}} \right\rangle } \right]} } }
\label {psi_k}
\end{equation}
\end{widetext}
are the eigenstates of the Hamiltonian ${\hat H^{MF}} = \hat H_{eff}^{MF} + {\hat H_{1ph}} + {\hat H_{2ph}}$, and using the roots corresponding to the minimum of the free energy $F =  - k_B T\ln Z$, where $Z = \sum\limits_k {e^{ - {{E_k } \mathord{\left/
 {\vphantom {{E_k } {k_B T}}} \right. \kern-\nulldelimiterspace} {k_B T}}} } $ is the partition function, various thermodynamic averages included in $\hat H_{eff}^{MF}$ can be calculated:
\begin{eqnarray}
\Delta _{ex,A\left( B \right)}^\sigma   = \sum\limits_k {\frac{{\left\langle {\psi _k \left| {X_{i_A \left( {i_B } \right)}^{s,\sigma } } \right|\psi _k } \right\rangle e^{{{ - E_k } \mathord{\left/
 {\vphantom {{ - E_k } {k_B T}}} \right.
 \kern-\nulldelimiterspace} {k_B T}}} }}{Z}}, \nonumber \\
m_{A\left( B \right)}  = \sum\limits_k {\frac{{\left\langle {\psi _k \left| {S_{i_{A\left( B \right)} }^z } \right|\psi _k } \right\rangle e^{{{ - E_k } \mathord{\left/
 {\vphantom {{ - E_k } {k_B T}}} \right.
 \kern-\nulldelimiterspace} {k_B T}}} }}{Z}}, \nonumber \\
n_{HS,A\left( B \right)}  = \sum\limits_k {\frac{{\left\langle {\psi _k \left| {\sum\limits_\sigma  {X_i^{\sigma ,\sigma } } } \right|\psi _k } \right\rangle e^{{{ - E_k } \mathord{\left/
 {\vphantom {{ - E_k } {k_B T}}} \right.
 \kern-\nulldelimiterspace} {k_B T}}} }}{Z}}\nonumber.
\end{eqnarray}
Thus, when solving Eq.~\ref{self_eig}, one deals with a self-consistent problem of finding the eigenstates and the eigenvalues of the effective Hamiltonian in the mean field approximation.

In Eq.~\ref{psi_k} $\left| {\sigma ,{n_{1ph}},{n_{2ph}}} \right\rangle $ and $\left| {s,{n_{1ph}},{n_{2ph}}} \right\rangle $ is the orthonormalized basis of functions in the form of the direct product of the spin projection operator eigenstates $\left| \sigma  \right\rangle $, $\sigma  =  \pm 1, 0$ in the case of the HS state and $\left| s \right\rangle $ for the LS state and the harmonic oscillators $\left| {{n_{1ph}}} \right\rangle  = \frac{1}{{\sqrt {{n_{1ph}}!} }}{\left( {{a^\dag }} \right)^{{n_{1ph}}}}\left| {0,0,...,0} \right\rangle $, $\left| {{n_{2ph}}} \right\rangle  = \frac{1}{{\sqrt {{n_{2ph}}!} }}{\left( {{b^\dag }} \right)^{{n_{2ph}}}}\left| {0,0,...,0} \right\rangle $. Here ${n_{1\left( 2 \right)ph}} = 0,1,2,...,{N_{1\left( 2 \right)ph}}$, where ${N_{1\left( 2 \right)ph}}$ is the number of phonons beginning with which the energy of the ground state $\left| \psi_0 \right\rangle$ is almost the same for ${n_{1\left( 2 \right)ph}} > {N_{1\left( 2 \right)ph}}$, i.e., ${E_0}\left( {{N_{1\left( 2 \right)ph}} + 1} \right) \simeq {E_0}\left( {{N_{1\left( 2 \right)ph}}} \right)$ (the error of the calculation is below 1\%). When considering various temperature effects, it is necessary to control the constancy of the energies $E_k$ of the lowest excited states $\left| \psi_k \right\rangle$. In other words, $N_{ph}$ is the number of phonons that should be taken into account at a given electron--vibration coupling in order to form a "phonon coat" of the ground and lowest excited states.

The terms describing the interactions beyond the mean field approximation have the form
\begin{eqnarray}
\delta {\hat H_{ex}} = \frac{1}{2}{J'_{ex}}\sum\limits_{\left\langle {i,j} \right\rangle } {\left[ {\delta \mbox{\boldmath{$\hat d$}}_i^\dag  \cdot \delta {\mbox{\boldmath{$\hat d$}}_j} + \delta {\mbox{\boldmath{$\hat d$}}_i} \cdot \delta \mbox{\boldmath{$\hat d$}}_j^\dag } \right]} - \nonumber \\
- \frac{1}{2}{J''_{ex}}\sum\limits_{\left\langle {i,j} \right\rangle } {\left[ {\delta \mbox{\boldmath{$\hat d$}}_i^\dag  \cdot \delta \mbox{\boldmath{$\hat d$}}_j^\dag  + \delta {\mbox{\boldmath{$\hat d$}}_i} \cdot \delta {\mbox{\boldmath{$\hat d$}}_j}} \right]};
\label {delta_H_ex}
\end{eqnarray}
\begin{equation}
\delta {\hat H_S} = \frac{1}{2}J\sum\limits_{\left\langle {i,j} \right\rangle } {\delta {\mbox{\boldmath{$\hat S$}}_i} \cdot \delta {\mbox{\boldmath{$\hat S$}}_j}}  = \frac{1}{2}J\sum\limits_{\xi ,\left\langle {i,j} \right\rangle } {\delta \hat S_i^\xi  \cdot \delta \hat S_j^\xi } ,
\label {delta_H_S}
\end{equation}
where $\xi = x, y, z;$
\begin{equation}
\delta {\hat H_{{n_{LS}}{n_{LS}}}} = \frac{1}{2}\tilde J\sum\limits_{\left\langle {i,j} \right\rangle } {\delta X_i^{s,s} \cdot \delta X_j^{s,s}}
\label {delta_H_nn}
\end{equation}

Below, we briefly present formulas for the generalized spin-wave (SW) approximation. For more detailed explanations, see the works~\cite{Nasu2016, PhysRevB.88.205110, Onufrieva}.
We introduce $\delta {\mathcal{\hat O}_{i,\xi }} = {\mathcal{\hat O}_{i,\xi }} - {\left\langle {{\mathcal{\hat O}_\xi }} \right\rangle _F}$, where $i$ belongs to the sublattice $F$. By $\mathcal{\hat O}_\xi$ we mean any of the $\mbox{\boldmath{$\hat d$}}$ and $\mbox{\boldmath{$\hat S$}}$ operators components. A set of the MFs $\left\{ {{{\left\langle {{\mathcal{\hat O}_\xi }} \right\rangle }_F}} \right\}$ is obtained in $\hat H^{MF}$ self-consistently in the numerical calculations. The fluctuation parts of the local operators are expanded by the Hubbard $X$-operators, which are defined by the eigenstates of $\hat H^{MF}$. This is given by
\begin{equation}
\delta {\mathcal{\hat O}_{i,\xi }} = \sum\limits_{m,n} {\left\langle {i,m\left| {{\delta \mathcal{\hat O}_{i,\xi }}} \right|i,n} \right\rangle X_i^{m,n}} ,
\label {O_xi_in_X}
\end{equation}
where $\left| {i,m} \right\rangle $ ($m = 0, 1, . . ., \mathcal{N}$) is the $m$th eigenstate of $\hat H^{MF}$ at site $i$, and the Hubbard $X$-operators are defined as $X_i^{m,n} = \left| {i,m} \right\rangle \left\langle {n,i} \right|$. Since the eigenenergy of $\left| {i,m} \right\rangle $ and the matrix elements ${\gamma _{{\mathcal{O}_{i,\xi }}}}\left( {m,n} \right) = {\left\langle {i,m\left| {\delta {\mathcal{\hat O}_{i,\xi }}} \right|i,n} \right\rangle }$ do not depend explicitly on $i$ but depend on the sublattice to which the site $i$ belongs, we denote them by $E_m^F$ and $\gamma _{{\mathcal{O}_\xi }}^F\left( {m,n} \right)$, respectively.

In the generalized Holstein-Primakoff (HP) representation
\begin{equation}
X_i^{m,0} = c_{im}^\dag {\left( {M - \sum\limits_{n = 1}^{\mathcal{N}} {c_{in}^\dag {c_{in}}} } \right)^{{1 \mathord{\left/
 {\vphantom {1 2}} \right.
 \kern-\nulldelimiterspace} 2}}},
\label {}
\end{equation}
and $X_i^{0,m} = {\left( {X_i^{m,0}} \right)^\dag }$ for $m \ge 1$,
\begin{equation}
X_i^{m,n} = c_{im}^\dag {c_{in}},
\label {}
\end{equation}
for $m,n \ge 1$, and
\begin{equation}
X_i^{0,0} = M - \sum\limits_{n = 1}^\mathcal{N} {c_{in}^\dag {c_{in}}},
\end{equation}
where $c_{in}^\dag $ (${c_{in}}$) is the creation (annihilation) operator of the HP boson at site $i$, and ${c_{in}}$ takes ${a_{in}}$ and ${b_{in}}$ when the site $i$ belongs to the sublattices $A$ and $B$, respectively. We
define $\mathcal{N}$ as the number of the excited states and $M \equiv X_i^{0,0} + \sum\limits_{n = 1}^{\mathcal{N}} {c_{in}^\dag {c_{in}}} $, where the constraint $M = 1$ is imposed at each site.

By expanding the Hubbard X-operators in terms of $1 / M$ and taking into account that for the ground state at a temperature close to zero we will have $\left\langle {i,0\left| {\delta {\mathcal{\hat O}_{i,\xi }}} \right|i,0} \right\rangle  = \left\langle {i,0\left| {{\mathcal{\hat O}_{i,\xi }}} \right|i,0} \right\rangle  - \left\langle {{\mathcal{\hat O}_{i,\xi }}} \right\rangle  \approx 0$ or ${\gamma _{{\mathcal{O}_\xi }}}\left( {0,0} \right) \approx 0$, the Hamiltonian $\hat H$ is expressed by the HP boson operators up to the quadratic order as
\begin{eqnarray}
\hat H_{SW} = \sum\limits_{\mbox{\scriptsize \boldmath{$q$}},m \ge 1} {\left( {\Delta E_m^Aa_{\mbox{\scriptsize \boldmath{$q$}}m}^\dag {a_{\mbox{\scriptsize \boldmath{$q$}}m}} + \Delta E_m^Bb_{\mbox{\scriptsize \boldmath{$q$}}m}^\dag {b_{\mbox{\scriptsize \boldmath{$q$}}m}}} \right)}  + \nonumber \\
+ \sum\limits_{\mbox{\scriptsize \boldmath{$q$}},m \ge ,n \ge 1} {{\gamma_{\mbox{\scriptsize \boldmath{$q$}}}}\left\{ {J_{mn}^{AB}a_{\mbox{\scriptsize \boldmath{$q$}}m}^\dag b_{ - \mbox{\scriptsize \boldmath{$q$}}n}^\dag  + \tilde J_{mn}^{AB}a_{\mbox{\scriptsize \boldmath{$q$}}m}^\dag {b_{\mbox{\scriptsize \boldmath{$q$}}n}} + {\rm{H}}{\rm{.c}}{\rm{.}}} \right\}}.
\label {H_SW_1}
\end{eqnarray}
We define $\Delta E_n^F = E_n^F - E_0^F$, and $\gamma_{\mbox{\scriptsize \boldmath{$q$}}} = \sum\limits_h {{e^{i\mbox{\scriptsize \boldmath{$q$}} \cdot \mbox{\scriptsize \boldmath{$h$}}}}} $, where $\mbox{\boldmath{$h$}}$ is the vector connecting NN sites. $c_{\mbox{\scriptsize \boldmath{$q$}}m} = \frac{1}{{\sqrt {{N \mathord{\left/
 {\vphantom {N 2}} \right.
 \kern-\nulldelimiterspace} 2}} }}\sum\limits_{i \in F} {{e^{ - i\mbox{\scriptsize \boldmath{$q$}} \cdot {\mbox{\scriptsize \boldmath{$r$}}_i}}}{c_{im}}} $. The hopping integrals of the HP bosons are given by
\begin{equation}
J_{mn}^{AB} = \sum\limits_{\mathcal{O},\xi } {{J_{{\mathcal{O}_\xi }}}\gamma _{{\mathcal{O}_\xi }}^A\left( {m,0} \right)\gamma _{{\mathcal{O}_\xi }}^B\left( {n,0} \right)},
\end{equation}
\begin{equation}
\tilde J_{mn}^{AB} = \sum\limits_{\mathcal{O},\xi } {{J_{{\mathcal{O}_\xi }}}\gamma _{{\mathcal{O}_\xi }}^A\left( {m,0} \right)\gamma _{{\mathcal{O}_\xi }}^{B * }\left( {n,0} \right)}.
\end{equation}
Using the Bogoliubov transformation~\cite{Colpa78}, Eq.~\ref{H_SW_1} is diagonalized as
\begin{equation}
\hat H_{SW} = \sum\limits_{\mbox{\scriptsize \boldmath{$q$}},\mu } {\left( {\omega _{\mbox{\scriptsize \boldmath{$q$}}\mu }^\alpha \alpha _{\mbox{\scriptsize \boldmath{$q$}}\mu }^\dag {\alpha _{\mbox{\scriptsize \boldmath{$q$}}\mu }} + \omega _{\mbox{\scriptsize \boldmath{$q$}}\mu }^\beta \beta _{\mbox{\scriptsize \boldmath{$q$}}\mu }^\dag {\beta _{\mbox{\scriptsize \boldmath{$q$}}\mu }}} \right)}  + {\rm{const}}
\label {H_SW_2},
\end{equation}
where ${\omega _{\mbox{\scriptsize \boldmath{$q$}}\mu }^\alpha }$ and ${\omega _{\mbox{\scriptsize \boldmath{$q$}}\mu }^\beta }$ are the energies for the bosons ${{\alpha _{\mbox{\scriptsize \boldmath{$q$}}\mu }}}$ and ${{\beta _{\mbox{\scriptsize \boldmath{$q$}}\mu }}}$, respectively.

Based on the generalized spin-wave method introduced above, we formulate the excitation spectra. The dynamical susceptibilities at zero temperature are given as
\begin{eqnarray}
{\chi _{\xi \xi '}}\left( {\mbox{\boldmath{$q$}},\omega } \right) &=& i\int\limits_0^\infty  {dt\left\langle {0\left| {\left[ {\delta {\mathcal{\hat O}_{\mbox{\scriptsize \boldmath{$q$}}\xi }}\left( t \right),\delta {\mathcal{\hat O}_{ - \mbox{\scriptsize \boldmath{$q$}}\xi '}}} \right]} \right|0} \right\rangle {e^{i\omega t - \delta t}}}  = \nonumber \\
&=& - \int\limits_{ - \infty }^\infty  {dE\frac{{{A_{\xi \xi '}}\left( {\mbox{\boldmath{$q$}},E} \right)}}{{\omega  - E + i \delta }}},
\end{eqnarray}
where $\left| 0 \right\rangle $ is the vacuum for ${{\alpha _{\mbox{\scriptsize \boldmath{$q$}}\mu }}}$ and ${{\beta _{\mbox{\scriptsize \boldmath{$q$}}\mu }}}$, ${{A_{\xi \xi '}}\left( {\mbox{\boldmath{$q$}},E} \right)}$ is the spectral function, $ \delta $ is an infinitesimal constant, and $\delta {\mathcal{\hat O}_{\mbox{\scriptsize \boldmath{$q$}}\xi }} = \frac{1}{N} \sum\limits_i {\delta {\mathcal{\hat O}_{i\xi }}{e^{ - i\mbox{\scriptsize \boldmath{$q$}} \cdot {\mbox{\scriptsize \boldmath{$r$}}_i}}}} $. The spectral function can be represented as
\begin{equation}
{A_{\xi \xi '}}\left( {\mbox{\boldmath{$q$}},E} \right) =  - \frac{1}{\pi }\sum\limits_{\mu \eta}  {W_{\mbox{\scriptsize \boldmath{$q$}}\xi \mu }^\eta W_{\mbox{\scriptsize \boldmath{$q$}}\xi '\mu }^{\eta  * }{\mathop{\rm Im}\nolimits} \left( {\frac{1}{{E - \omega  + i \delta }}} \right)},
\label {Spectral_function}
\end{equation}
here $W_{\mbox{\scriptsize \boldmath{$q$}}\xi \mu }^\eta  = \sum\limits_{\mbox{\scriptsize \boldmath{$q$}}'} {\left\langle {0\left| {\delta {\mathcal{\hat O}_{\xi \mbox{\scriptsize \boldmath{$q$}}}}} \right|\mu ,\mbox{\boldmath{$q$}}',\eta } \right\rangle \left\langle {\mu ,\mbox{\boldmath{$q$}}',\eta \left| {\delta {\mathcal{\hat O}_{\xi ' - \mbox{\scriptsize \boldmath{$q$}}}}} \right|0} \right\rangle } $, where $\left| {\mu ,\mbox{\boldmath{$q$}},\eta } \right\rangle  = \eta _{\mbox{\scriptsize \boldmath{$q$}}\mu }^\dag \left| 0 \right\rangle $ and ${\eta _{\mbox{\scriptsize \boldmath{$q$}}\mu }} = \left( {{\alpha _{\mbox{\scriptsize \boldmath{$q$}}\mu }},{\beta _{\mbox{\scriptsize \boldmath{$q$}}\mu }}} \right)$ for $\eta  = \left( {\alpha ,\beta } \right)$.

\begin{figure}
\includegraphics[width=9.0cm, height=4.0cm]{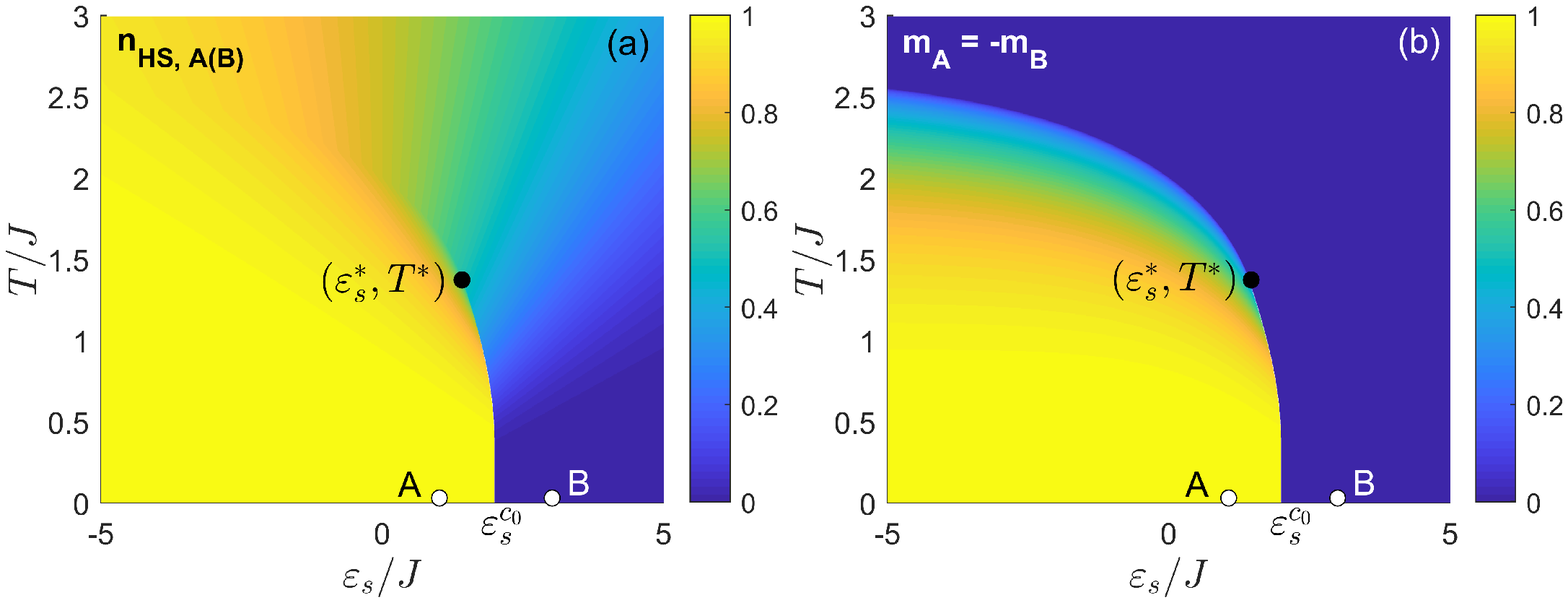}
\caption{\label{fig:1} Phase diagrams of the HS state population (a) and the magnetization (b) for two sublattices at ${J''_{ex}} = 0$. $\left( {{T^ * },\varepsilon _S^ * } \right)$ is the tricritical point. At points "A" (${{{\varepsilon _S}} \mathord{\left/
 {\vphantom {{{\varepsilon _S}} J}} \right.
 \kern-\nulldelimiterspace} J} = 1$,~$T = 0$~K) and "B" (${{{\varepsilon _S}} \mathord{\left/
 {\vphantom {{{\varepsilon _S}} J}} \right.
 \kern-\nulldelimiterspace} J} = 3$,~$T = 0$~K) the collective excitations spectrum is calculated (Fig.~\ref{fig:2}). Calculations were performed at $J = 28$~K, ${J''_{ex}} = 0$, $g_1 = 0$, and $g_2 = 0$.}
\end{figure}

To begin with, as an example, consider the simplest case. Figure~\ref{fig:1} shows the calculated phase diagrams of the magnetization and population of the HS state for the two sublattices $A$ and $B$ at ${J''_{ex}} = 0$, $g_1 = 0$, and $g_2 = 0$ in the coordinates temperature $T$ -- spin gap ${\varepsilon _S}$. The calculation was performed at $J = 28$~K~\cite{PhysRevB.79.214421}. Hereinafter, the temperature and spin gap are given in terms of exchange parameter $J$. It can be seen that, due to the presence of the cooperative exchange coupling $J$, the ground antiferromagnetically ordered HS-AFM state is preserved in a system up to ${\varepsilon _S} = \varepsilon _S^{{C_0}} \approx 2J$ [Fig.~\ref{fig:1}], although in the single-ion picture at ${\varepsilon _S} > 0$, the LS state is a ground state. The growth of the critical $\varepsilon _S^{{C_0}}$ value by the expense of the cooperative effects is quite clear, since the exchange coupling stabilizes the HS state via lowering its energy. At ${\varepsilon _S} > \varepsilon _S^{{C_0}}$, the ground HS-AFM state changes for the nonmagnetic LS state [Fig.~\ref{fig:1}]. In the range of ${\varepsilon _S} < \varepsilon _S^{{C_0}}$ [Fig.~\ref{fig:1}], with increasing temperature, the system undergoes a second-order phase transition from the HS-AFM state to the paramagnetic state at ${\varepsilon _S} < \varepsilon _S^ * $ and a first-order transition at $\varepsilon _S^ *  < {\varepsilon _S} < \varepsilon _S^{{C_0}}$. In the diagram, one can clearly see the existence of a tricritical point [$\left( {{T^ * },\varepsilon _S^ * } \right)$ in Fig.~\ref{fig:1}], at which the line of the second-order phase transitions continuously passes to the line of the first-order ones.

\begin{figure}
\includegraphics[width=8.5cm, height=11.0cm]{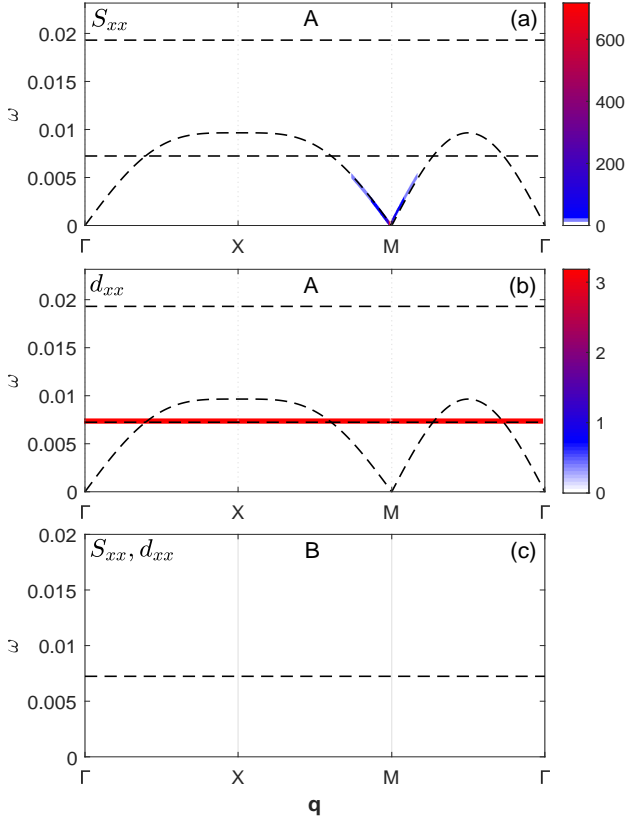}
\caption{\label{fig:2} Contour maps of the dynamical spin $\mbox{\boldmath{$S$}}$ and $\mbox{\boldmath{$d$}}$ correlation functions Eq.~\ref{Spectral_function} at points ``A'' [(a), (b)] and ``B'' (c) on phase diagrams in Fig.~\ref{fig:1}. The dashed black lines represent the dispersion relations of the collective modes.}
\end{figure}

The dashed line in Fig.~\ref{fig:2} shows the collective excitations spectrum Eq.~\ref{H_SW_2} calculated at points ``A'' and ``B'' on the phase diagrams in Fig.~\ref{fig:1}. The color shows the spectral weight distribution function Eq.~\ref{Spectral_function}. In what follows, we will be interested only in the spectrum itself, without the distribution of the spectral weight.

Figure~\ref{fig:2} clearly shows the presence of ordinary spin-wave (magnon) excitations, as in the Heisenberg model for spin $S=1$ and dispersionless excitations due to the presence of a singlet $\left| s \right\rangle $ in addition to the triplet $\left| \sigma  \right\rangle $ state.

\begin{figure}
\includegraphics[width=9.0cm, height=7.0cm]{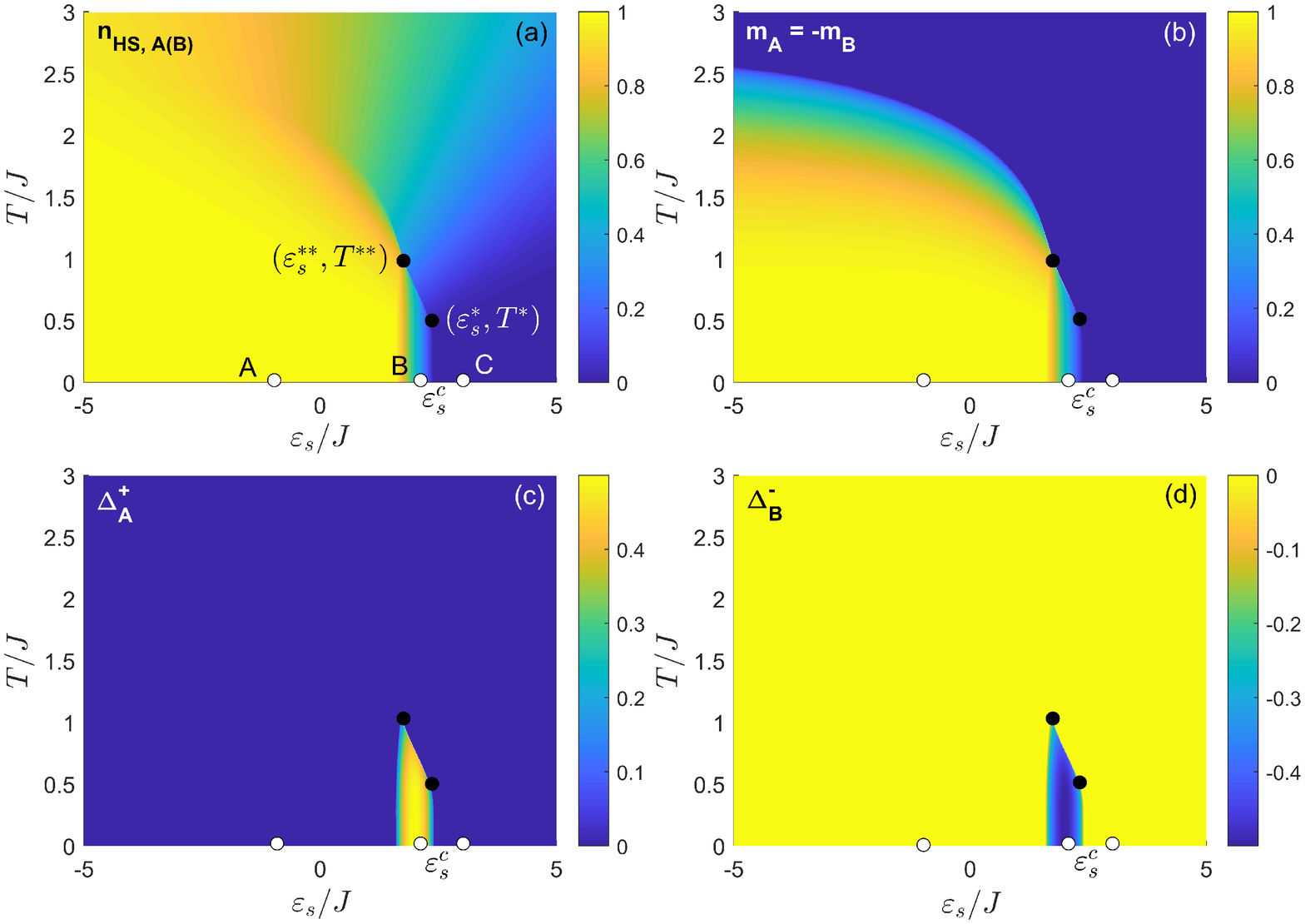}
\caption{\label{fig:3} Calculated phase diagrams of (a) population $n_{HS}$ of the HS state, (b) magnetization $m$, (c), (d) exciton order parameter components for sublattices $A$ and $B$. $\left( {{T^ * },\varepsilon _S^ * } \right)$ is the tricritical point, $\left( {T^{**} ,\varepsilon _S^{**} } \right)$ is the bicritical point. At points ``A'' (${{{\varepsilon _S}} \mathord{\left/
 {\vphantom {{{\varepsilon _S}} J}} \right.
 \kern-\nulldelimiterspace} J} =  - 1$,~$T = 0$~K), ``B'' (${{{\varepsilon _S}} \mathord{\left/
 {\vphantom {{{\varepsilon _S}} J}} \right.
 \kern-\nulldelimiterspace} J} = 2.1$,~$T = 0$~K), and ``C'' (${{{\varepsilon _S}} \mathord{\left/
 {\vphantom {{{\varepsilon _S}} J}} \right.
 \kern-\nulldelimiterspace} J} = 3$,~$T = 0$~K) the collective excitations spectrum is calculated (Fig.~\ref{fig:4}). Calculations were performed at $J = 28$~K, ${J''_{ex}} = 0.6J$, $g_1 = 0$, and $g_2 = 0$.}
\end{figure}

At ${J''_{ex}} = 0.6J$, the phase diagram becomes richer [Fig.~\ref{fig:3}]. Figure~\ref{fig:3} shows the calculated phase diagrams of population $n_{HS}$ of the HS state (a), magnetization $m$ (b), exciton order parameter components ${\Delta ^{{ +  \mathord{\left/
 {\vphantom { +   - }} \right.
 \kern-\nulldelimiterspace}  - }}}$ [Figs.~\ref{fig:3}(c) and~\ref{fig:3}(d)] for two sublattices $A$ and $B$ in the coordinates temperature $T$ -- spin gap ${\varepsilon _S}$. Electron-phonon interaction constants are still equal to zero, $g_{1\left( 2 \right)} = 0$. It can be seen that $n_{HS,A} = n_{HS,B}$ (a); $m_A = -m_B$, the long-range antiferromagnetic order is implemented in a system (b); $\Delta _A^ +  =  - \Delta _B^ - $ [Figs.~\ref{fig:3}(c) and~\ref{fig:3}(d)], in this case, we have $\Delta _A^ -  = \Delta _B^ +  = 0$, $\Delta _{A\left( B \right)}^0 = 0$.

Near $\varepsilon _S^C$, the exciton condensate region arises, which coexists with the long-range antiferromagnetic order. Moreover, the comparison of Fig.~\ref{fig:3} and Fig.~\ref{fig:1} shows that the formation of the exciton condensate facilitates the antiferromagnetic ordering and the occurrence of the magnetization in the range of parameters $T$ and ${\varepsilon _S}$ where there was no magnetic order at ${J''_{ex}} = 0$ [Fig.~\ref{fig:3}(b), $\varepsilon _S^C > \varepsilon _S^{{C_0}}$]. Physically, this is quite clear from the structure of the exciton order parameter. In particular, at $\Delta _A^ -  \ne 0$, we have $\Delta _A^ +  = 0$; then, $\Delta _B^ +  =  - \Delta _A^ - $ and $\Delta _B^ -  = 0$. On the contrary, $\Delta _A^ +  \ne 0$, we have $\Delta _A^ -  = 0$; then, $\Delta _B^ -  =  - \Delta _A^ + $ and $\Delta _B^ +  = 0$. The nonzero corresponding values ${\Delta ^{{ +  \mathord{\left/
 {\vphantom { +   - }} \right.
 \kern-\nulldelimiterspace}  - }}}$ on different sublattices allow the coexistence of an exciton condensate with antiferromagnetism and facilitates its formation (for more details, see~\cite{Orlov_PhysRevB.104.195103}).

Along with the tricritical point (${T^ * }$ and $\varepsilon _S^ * $ in Fig.~\ref{fig:1} and Fig.~\ref{fig:3}), the phase diagram contains a bicritical point in (${T^{ *  * }}$ and $\varepsilon _S^{ *  * }$ in Fig.~\ref{fig:3}), at which the second-order phase transition line splits into two second- and first-order phase transition lines (Fig.~\ref{fig:3}), according to the Gibbs phase rule~\cite{Orlov_PhysRevB.104.195103}.

\begin{figure}
\includegraphics[width=8.5cm, height=11.0cm]{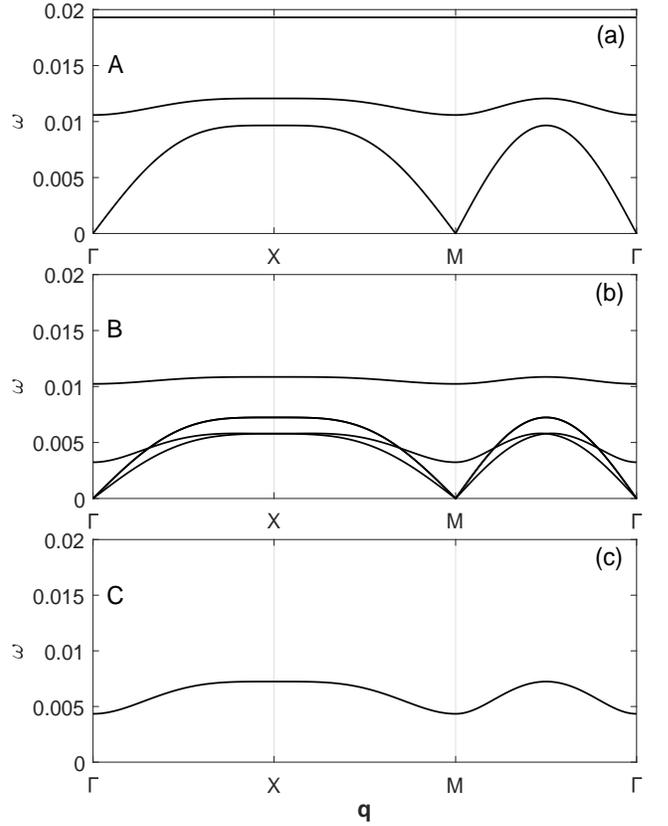}
\caption{\label{fig:4} The collective excitations spectrum Eq.~\ref{H_SW_2} calculated at points ``A'', ``B'' and ``C'' on the phase diagrams in Fig.~\ref{fig:3}.}
\end{figure}

The solid line in Fig.~\ref{fig:4} shows the collective excitations spectrum Eq.~\ref{H_SW_2} calculated at points ``A'', ``B'' and ``C'' on the phase diagrams in Fig.~\ref{fig:3}.

In the ordered region [Figs.~\ref{fig:4}(a) and \ref{fig:4}(b)] the same as in the case Figs.~\ref{fig:2}(a) and \ref{fig:2}(b), the spectrum is gapless with a linear dispersion law for small wave vector values. This corresponds to the massless Goldstone mode appearance upon spontaneous symmetry breaking as a result of the phase transition.

The inclusion of the diagonal electron-phonon interaction $g_1$ does not lead to qualitatively new results. The region of the exciton condensate [Figs.~\ref{fig:3}(c) and \ref{fig:3}(d)] decreases, while the spectrum of collective excitations [Fig.\ref{fig:4}] does not change.

\begin{figure}
\includegraphics[width=9.0cm, height=7.0cm]{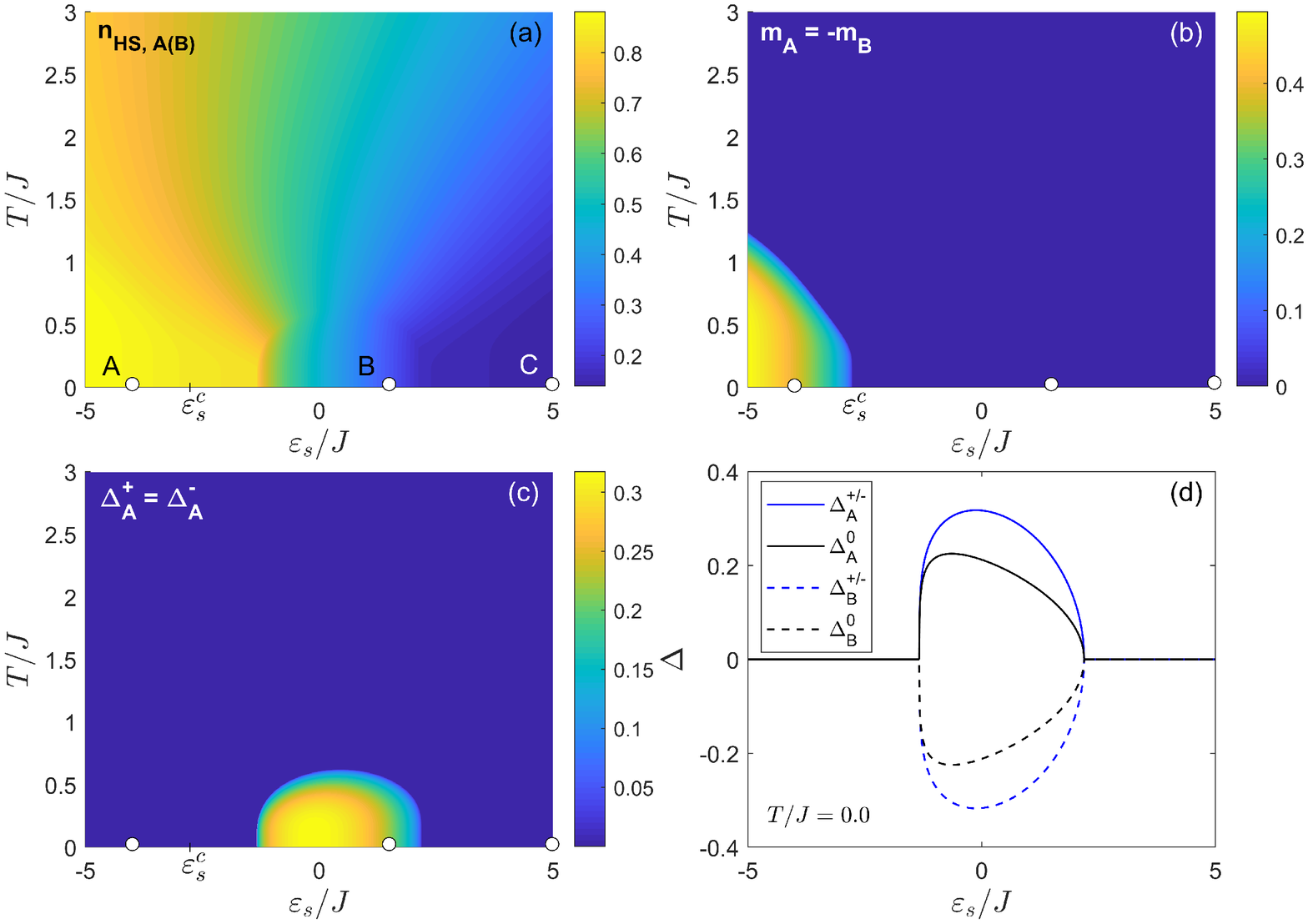}
\caption{\label{fig:5} Calculated phase diagrams of (a) population $n_{HS}$ of the HS state, (b) magnetization $m$, (c) exciton order parameter components for sublattices A and B. (d) The dependence of the exciton order parameter components on the spin gap at $T=0$. At points ``A'' (${{{\varepsilon _S}} \mathord{\left/
 {\vphantom {{{\varepsilon _S}} J}} \right.
 \kern-\nulldelimiterspace} J} =  - 4$), ``B'' (${{{\varepsilon _S}} \mathord{\left/
 {\vphantom {{{\varepsilon _S}} J}} \right.
 \kern-\nulldelimiterspace} J} = 1.5$), and ``C'' (${{{\varepsilon _S}} \mathord{\left/
 {\vphantom {{{\varepsilon _S}} J}} \right.
 \kern-\nulldelimiterspace} J} = 5$) the collective excitations spectrum is calculated (Fig.~\ref{fig:6}). Calculations were performed at $J = 28$~K, ${J''_{ex}} = 0.6J$, $g_1 = 0$, and $g_2 = 0.014$~eV.}
\end{figure}

In contrast to the diagonal one, the presence of an off-diagonal electron-phonon interaction $g_2$ leads to significant changes. It can be seen in Fig.~\ref{fig:5}(b) that the region of antiferromagnetic order decreases in comparison with Figs.~\ref{fig:1}(b) and \ref{fig:3}(b). The $\varepsilon _S^C$ critical value is greatly reduced [Fig.~\ref{fig:5}(b), $\varepsilon _S^C < 0$]. The exciton condensate phase still forms near ${\varepsilon _S} = 0$, but does not coexist with antiferromagnetism (these phases are different). This is explained by the structure of the exciton order parameter [Fig.~\ref{fig:5}(c)]. In contrast to the case discussed above, now $\Delta _{A\left( B \right)}^ +  = \Delta _{A\left( B \right)}^ - $, $\Delta _{A\left( B \right)}^0 \ne 0$

\begin{figure}
\includegraphics[width=8.5cm, height=11.0cm]{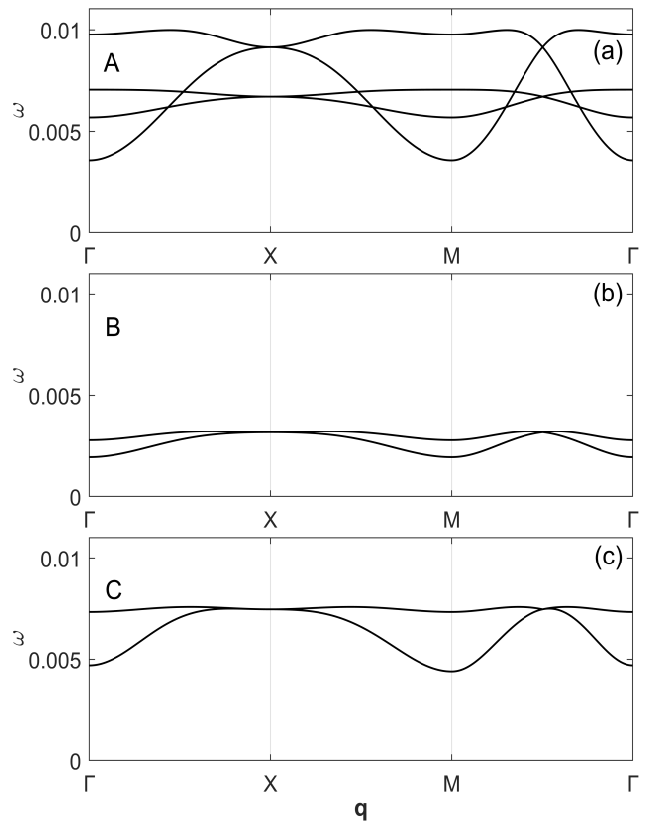}
\caption{\label{fig:6} Low lying the collective excitations spectrum Eq.~\ref{H_SW_2} calculated at points ``A'', ``B'' and ``C'' on the phase diagrams in Fig.~\ref{fig:5}.}
\end{figure}

The off-diagonal electron-phonon interaction $g_2$ leads to the opening of a gap in the collective excitations spectrum. The Fig.~\ref{fig:6} shows only the low-lying part of the spectrum, calculated at points ``A'', ``B'' and ``C'' on the phase diagrams in Fig.~\ref{fig:5}.

\section{\label{sec:4}Photoinduced excitonic order enhancement and magnetic transitions}

Next, we consider the excitation with a laser pulse and discuss how these different properties of the collective modes show up out of equilibrium, and under which conditions the order can be enhanced.

To study photoexcitation processes, it is necessary to take into account the interaction with external radiation. We write the complete Hamiltonian in the form
\begin {equation}
\hat H\left( t \right) = \hat H + \hat V\left( t \right),
\label {Ht_1}
\end {equation}
where $\hat V\left( t \right) = {d_0}E\left( t \right)\sum\limits_i {\sum\limits_{\sigma  =  - 1}^{ + 1} {\left( {X_i^{s,\sigma } + X_i^{\sigma ,s}} \right)} } $ is the interaction term. $E\left( t \right)$ denotes the external laser field, and we assume that the laser couples to the system through dipolar transitions (${d_0}$ is the effective dipolar moment). We prepare the equilibrium state at $T = 0$ at $t = 0$ and choose $E\left( t \right) = {E_{0}}\sin \left( {\Omega t} \right)\exp \left[ { - {{{{\left( {t - {t_p}} \right)}^2}} \mathord{\left/
 {\vphantom {{{{\left( {t - {t_p}} \right)}^2}} {\left( {2\sigma _p^2} \right)}}} \right.
 \kern-\nulldelimiterspace} {\left( {2\sigma _p^2} \right)}}} \right]$ with $\Omega = 5\tau_0^{-1}$, $\sigma_p = 3\tau_0$, $t_p = 15\tau_0$, and ${d_0}{E_0} = 0.12J$, where $\tau_0 = 10^{-12}$~sec.

In the MF approximation, the Eq.~\ref{Ht_1} takes the form
\begin {equation}
\hat H\left( t \right) = {\hat H_0} + \hat V\left( t \right),
\label {Ht_2}
\end {equation}
where ${\hat H_0} = \sum\limits_k {{E_k}\left| {{\psi _k}} \right\rangle \left\langle {{\psi _k}} \right|} $.
We consider the dynamics of the system in terms of the density matrix $\rho^0$ using the generalized master equation~\cite{Blum}
\begin {eqnarray}
\dot \rho _{kl}^0\left( t \right) =  - i{\omega _{kl}}\rho _{kl}^0\left( t \right) - \sum\limits_{mn} {\rho _{mn}^0{R_{klmn}}}  - \nonumber \\
- \frac{i}{\hbar }\sum\limits_n {\left[ {\left\langle {{\psi _k}\left| {\hat V\left( t \right)} \right|{\psi _n}} \right\rangle \rho _{nl}^0\left( t \right) - \rho _{kn}^0\left( t \right)\left\langle {{\psi _n}\left| {\hat V\left( t \right)} \right|{\psi _l}} \right\rangle } \right]}.
\label {Master_1}
\end {eqnarray}

\begin{figure*}
\includegraphics[width=16.0cm, height=11.0cm]{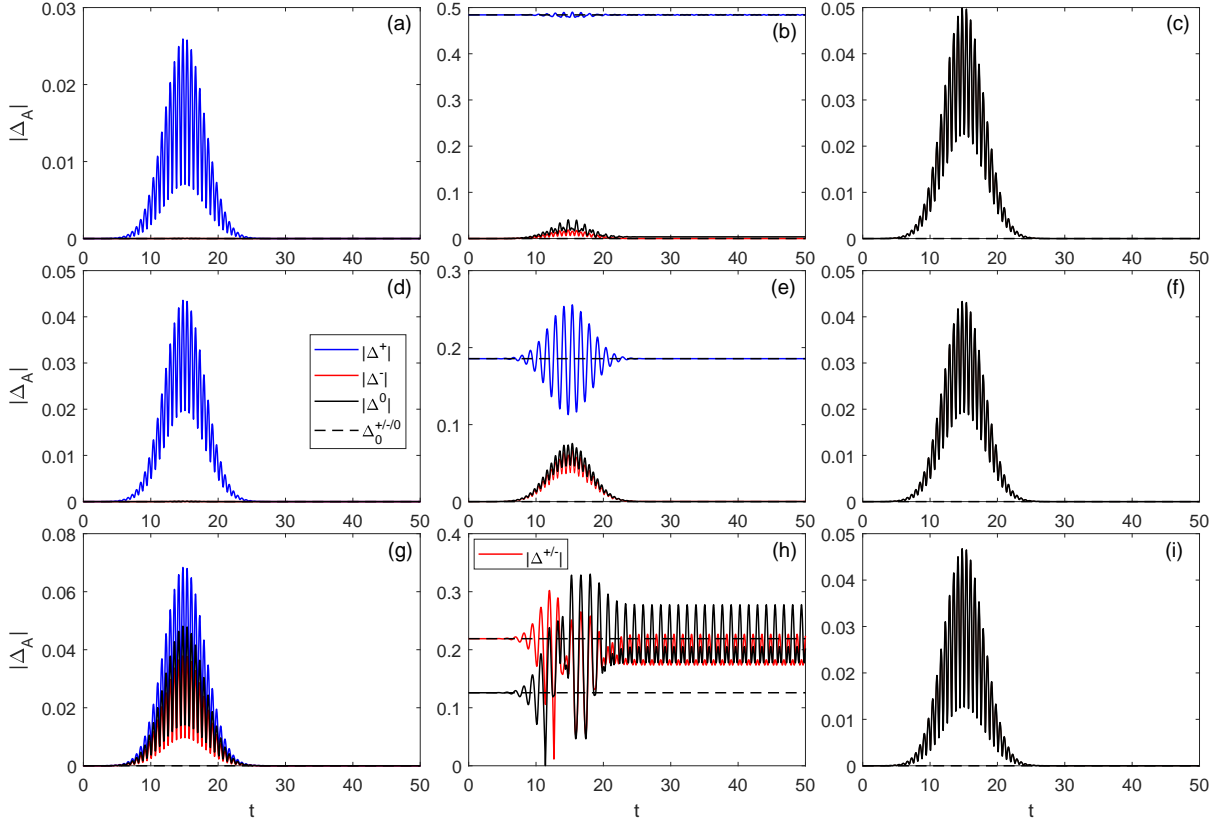}
\caption{\label{fig:7} Pulse radiation action result on the electronic subsystem. The exciton order parameter components time evolution (modulo value) for $A$ sublattice (a similar dependence takes place for sublattice $B$). The top row shows the dynamics calculation's results at points ``A'', ``B'' and ``C'' of the phase diagram in Fig.~\ref{fig:3} in the absence of electron-phonon interaction [Figs.~(a), (b), and (c), respectively]. Similarly, for the case of diagonal electron-phonon interaction $g_1 = 0.01$~eV [middle row, Figs. (d), (e), and (f)]. The bottom row shows the dynamics calculating results at points ``A'', ``B'' and ``C'' of the phase diagram in Fig.~\ref{fig:5} in the presence of an off-diagonal electron-phonon interaction $g_2$ [Figs. (g), (h), and (i), respectively]. The red and blue lines are the same in the Fig. (h). Time $t$ is in units of $\tau_0 = 10^{-12}$~sec.}
\end{figure*}

\noindent The first term in Eq.~\ref{Master_1} describes the reversible motion in terms of the transition frequencies ${\omega _{kl}} = {{\left( {{E_k} - {E_l}} \right)} \mathord{\left/
 {\vphantom {{\left( {{E_k} - {E_l}} \right)} \hbar }} \right.
 \kern-\nulldelimiterspace} \hbar }$ between energy levels in the spin-crossover system, the second term describes relaxation, and the last term describes interaction with radiation. Damping processes are not we will not considered below ($R_{klmn} = 0$). If we introduce ${\Theta _{kn}}\left( t \right) = \frac{{{d_0}E\left( t \right)}}{\hbar }\left\langle {{\psi _k}\left| {\sum\limits_{\sigma  =  - 1}^{ + 1} {\left( {{X^{s,\sigma }} + {X^{\sigma ,s}}} \right)} } \right|{\psi _n}} \right\rangle $, then Eq.~\ref{Master_1} takes the form
\begin {eqnarray}
\dot \rho _{kl}^0\left( t \right) &=& - i{\omega _{kl}}\rho _{kl}^0\left( t \right) - \nonumber \\
 &-& i\sum\limits_n {\left[ {{\Theta _{kn}}\left( t \right)\rho _{nl}^0\left( t \right) - \rho _{kn}^0\left( t \right){\Theta _{nl}}\left( t \right)} \right]}.
\label {Master_2}
\end {eqnarray}

\begin{figure*}
\includegraphics[width=16.0cm, height=11.0cm]{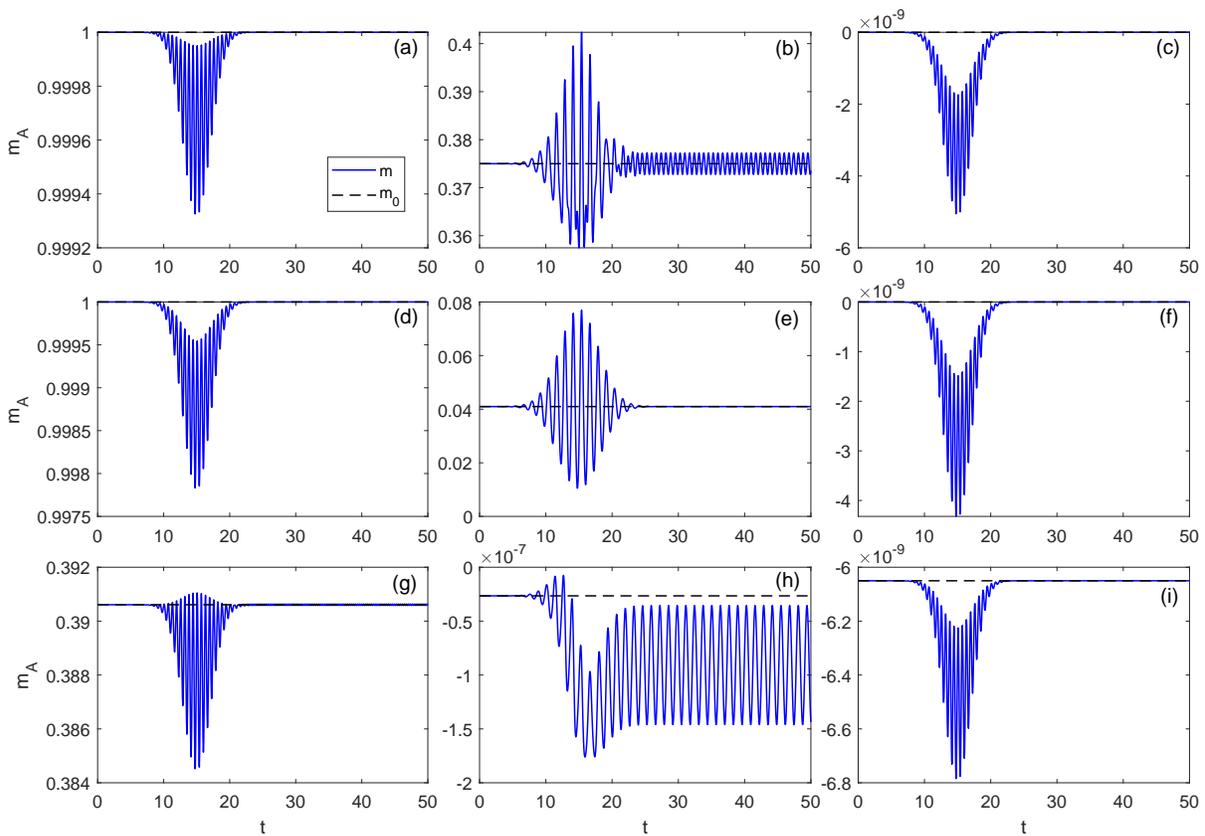}
\caption{\label{fig:8} Pulse radiation action result on magnetic subsystem. The sublattice magnetization ($m_A = - m_B$) calculation's results in a one-to-one correspondence for the cases considered in Fig.~\ref{fig:7}. Time $t$ is in units of $\tau_0 = 10^{-12}$~sec.}
\end{figure*}

Figs.~\ref{fig:7} and~\ref{fig:8} show the pulse radiation's action result on the electronic and magnetic subsystems, respectively. Fig.~\ref{fig:7} shows the exciton order parameter components time evolution (modulo value) for $A$ sublattice (a similar dependence takes place for sublattice $B$), obtained by solving the Eq.~\ref{Master_2}. The top row shows the dynamics calculating results at points ``A'', ``B'' and ``C'' of the phase diagram in Fig.~\ref{fig:3} in the absence of electron-phonon interaction [Figs.~(a), (b), and (c), respectively]. It can be seen that after turning off the external radiation, the system returns to the order parameter components equilibrium values shown by the dashed line. Turning on the diagonal electron-phonon interaction $g_1$ (middle row in Fig.~\ref{fig:7}) does not lead to a fundamental difference; after turning off the external action, the system also returns to its initial state [Figs. (d), (e), and (f)]. A fundamentally different situation arises in the presence of an off-diagonal electron-phonon interaction $g_2$. The bottom row in Fig.~\ref{fig:7} shows the dynamics calculation's results at points ``A'', ``B'' and ``C'' of the phase diagram in Fig.~\ref{fig:5} in the presence of an off-diagonal electron-phonon interaction $g_2$ [Figs. (g), (h), and (i), respectively]. In this case, after switching off the external perturbation, the order parameter components oscillations around a certain average position, which is different from the equilibrium one, are observed in the exciton phase [Fig.~\ref{fig:7}~(h)]. Thus, we can conclude that the exciton condensate is photoenhanced.

Figure~\ref{fig:8} presents the sublattice magnetization ($m_A = - m_B$) calculation's results in a one-to-one correspondence for the cases considered in Fig.~\ref{fig:7}. Just as in the case of Fig.~\ref{fig:7}(h), after perturbation turning off, the magnetization oscillations around a certain average position, different from the equilibrium one, shown by the dashed line, are seen in Fig.~\ref{fig:8}(h). Thus, one can conclude the presence of a photoinduced magnetic transition.

\section{\label{sec:5}Discussion}

In our previous work~\cite{Exciton_arxiv}, by finite-temperature Green's function method, we determined the quasiparticle (one-particle) exciton excitations spectrum of a Bose particles system described by the effective Hamiltonian~\ref{H_ex} outside the exciton condensate phase and showed its instability towards the formation of the excitonic Bose-condensate. Everywhere outside the exciton condensate phase, the spectrum has a gap that vanishes at the boundary of the second-order phase transition, which agrees with the results of this work and with the general idea that, below the phase transition critical point, a gapless Goldstone mode should arise, describing collective excitations in the exciton condensate phase. The off-diagonal electron-phonon interaction (in contrast to the diagonal interaction) leads to the gap opening in the spectrum of individual exciton excitations at the boundary of the second-order phase transition~\cite{Exciton_arxiv}. Its also agrees with the results of this work, where it is shown that the collective Goldstone mode in the exciton condensate phase acquires mass, due to off-diagonal electron-phonon interaction. In this case, the Bose spectrum of excitations (one-particle excitonic and collective in the excitonic phase) has a gap on both sides of the phase transition critical point. Thus, we can conclude that the gap presence or absence in the spectrum of quasiparticle excitations at the phase transition critical point corresponds to the gap presence or absence in the spectrum of collective excitations, and vice versa. The fact that the collective excitation spectrum has a gap plays an important role in the photoenhancement of the exciton condensate and may provide a new strategy for increasing the order parameter in similar systems (e.g., superconductors).

\begin{acknowledgments}
The study was supported by the Russian Science Foundation grant No. 22-22-20007, Krasnoyarsk Regional Science Foundation.
\end{acknowledgments}

\bibliography{manuscriptbibl}% Produces the bibliography via BibTeX.

\begin{thebibliography}{34}%
\makeatletter
\providecommand \@ifxundefined [1]{%
 \@ifx{#1\undefined}
}%
\providecommand \@ifnum [1]{%
 \ifnum #1\expandafter \@firstoftwo
 \else \expandafter \@secondoftwo
 \fi
}%
\providecommand \@ifx [1]{%
 \ifx #1\expandafter \@firstoftwo
 \else \expandafter \@secondoftwo
 \fi
}%
\providecommand \natexlab [1]{#1}%
\providecommand \enquote  [1]{``#1''}%
\providecommand \bibnamefont  [1]{#1}%
\providecommand \bibfnamefont [1]{#1}%
\providecommand \citenamefont [1]{#1}%
\providecommand \href@noop [0]{\@secondoftwo}%
\providecommand \href [0]{\begingroup \@sanitize@url \@href}%
\providecommand \@href[1]{\@@startlink{#1}\@@href}%
\providecommand \@@href[1]{\endgroup#1\@@endlink}%
\providecommand \@sanitize@url [0]{\catcode `\\12\catcode `\$12\catcode
  `\&12\catcode `\#12\catcode `\^12\catcode `\_12\catcode `\%12\relax}%
\providecommand \@@startlink[1]{}%
\providecommand \@@endlink[0]{}%
\providecommand \url  [0]{\begingroup\@sanitize@url \@url }%
\providecommand \@url [1]{\endgroup\@href {#1}{\urlprefix }}%
\providecommand \urlprefix  [0]{URL }%
\providecommand \Eprint [0]{\href }%
\providecommand \doibase [0]{https://doi.org/}%
\providecommand \selectlanguage [0]{\@gobble}%
\providecommand \bibinfo  [0]{\@secondoftwo}%
\providecommand \bibfield  [0]{\@secondoftwo}%
\providecommand \translation [1]{[#1]}%
\providecommand \BibitemOpen [0]{}%
\providecommand \bibitemStop [0]{}%
\providecommand \bibitemNoStop [0]{.\EOS\space}%
\providecommand \EOS [0]{\spacefactor3000\relax}%
\providecommand \BibitemShut  [1]{\csname bibitem#1\endcsname}%
\let\auto@bib@innerbib\@empty
%</preamble>
\bibitem [{\citenamefont {Nasu}\ \emph {et~al.}(2016)\citenamefont {Nasu},
  \citenamefont {Watanabe}, \citenamefont {Naka},\ and\ \citenamefont
  {Ishihara}}]{Nasu2016}%
  \BibitemOpen
  \bibfield  {author} {\bibinfo {author} {\bibfnamefont {J.}~\bibnamefont
  {Nasu}}, \bibinfo {author} {\bibfnamefont {T.}~\bibnamefont {Watanabe}},
  \bibinfo {author} {\bibfnamefont {M.}~\bibnamefont {Naka}},\ and\ \bibinfo
  {author} {\bibfnamefont {S.}~\bibnamefont {Ishihara}},\ }\bibfield  {title}
  {\bibinfo {title} {Phase diagram and collective excitations in an excitonic
  insulator from an orbital physics viewpoint},\ }\href
  {https://doi.org/10.1103/PhysRevB.93.205136} {\bibfield  {journal} {\bibinfo
  {journal} {Phys. Rev. B}\ }\textbf {\bibinfo {volume} {93}},\ \bibinfo
  {pages} {205136} (\bibinfo {year} {2016})}\BibitemShut {NoStop}%
\bibitem [{\citenamefont {Fausti}\ \emph {et~al.}(2011)\citenamefont {Fausti},
  \citenamefont {Tobey}, \citenamefont {Dean}, \citenamefont {Kaiser},
  \citenamefont {Dienst}, \citenamefont {Hoffmann}, \citenamefont {Pyon},
  \citenamefont {Takayama}, \citenamefont {Takagi},\ and\ \citenamefont
  {Cavalleri}}]{Science.331.2011}%
  \BibitemOpen
  \bibfield  {author} {\bibinfo {author} {\bibfnamefont {D.}~\bibnamefont
  {Fausti}}, \bibinfo {author} {\bibfnamefont {R.~I.}\ \bibnamefont {Tobey}},
  \bibinfo {author} {\bibfnamefont {N.}~\bibnamefont {Dean}}, \bibinfo {author}
  {\bibfnamefont {S.}~\bibnamefont {Kaiser}}, \bibinfo {author} {\bibfnamefont
  {A.}~\bibnamefont {Dienst}}, \bibinfo {author} {\bibfnamefont {M.~C.}\
  \bibnamefont {Hoffmann}}, \bibinfo {author} {\bibfnamefont {S.}~\bibnamefont
  {Pyon}}, \bibinfo {author} {\bibfnamefont {T.}~\bibnamefont {Takayama}},
  \bibinfo {author} {\bibfnamefont {H.}~\bibnamefont {Takagi}},\ and\ \bibinfo
  {author} {\bibfnamefont {A.}~\bibnamefont {Cavalleri}},\ }\bibfield  {title}
  {\bibinfo {title} {Light-induced superconductivity in a stripe-ordered
  cuprate},\ }\href {https://doi.org/10.1126/science.1197294} {\bibfield
  {journal} {\bibinfo  {journal} {Science}\ }\textbf {\bibinfo {volume}
  {331}},\ \bibinfo {pages} {189} (\bibinfo {year} {2011})},\ \Eprint
  {https://arxiv.org/abs/https://www.science.org/doi/pdf/10.1126/science.1197294}
  {https://www.science.org/doi/pdf/10.1126/science.1197294} \BibitemShut
  {NoStop}%
\bibitem [{\citenamefont {Kaiser}\ \emph {et~al.}(2014)\citenamefont {Kaiser},
  \citenamefont {Hunt}, \citenamefont {Nicoletti}, \citenamefont {Hu},
  \citenamefont {Gierz}, \citenamefont {Liu}, \citenamefont {Le~Tacon},
  \citenamefont {Loew}, \citenamefont {Haug}, \citenamefont {Keimer},\ and\
  \citenamefont {Cavalleri}}]{PhysRevB.89.184516}%
  \BibitemOpen
  \bibfield  {author} {\bibinfo {author} {\bibfnamefont {S.}~\bibnamefont
  {Kaiser}}, \bibinfo {author} {\bibfnamefont {C.~R.}\ \bibnamefont {Hunt}},
  \bibinfo {author} {\bibfnamefont {D.}~\bibnamefont {Nicoletti}}, \bibinfo
  {author} {\bibfnamefont {W.}~\bibnamefont {Hu}}, \bibinfo {author}
  {\bibfnamefont {I.}~\bibnamefont {Gierz}}, \bibinfo {author} {\bibfnamefont
  {H.~Y.}\ \bibnamefont {Liu}}, \bibinfo {author} {\bibfnamefont
  {M.}~\bibnamefont {Le~Tacon}}, \bibinfo {author} {\bibfnamefont
  {T.}~\bibnamefont {Loew}}, \bibinfo {author} {\bibfnamefont {D.}~\bibnamefont
  {Haug}}, \bibinfo {author} {\bibfnamefont {B.}~\bibnamefont {Keimer}},\ and\
  \bibinfo {author} {\bibfnamefont {A.}~\bibnamefont {Cavalleri}},\ }\bibfield
  {title} {\bibinfo {title} {Optically induced coherent transport far above
  ${T}_{c}$ in underdoped
  ${\mathrm{yba}}_{2}{\mathrm{cu}}_{3}{\mathrm{o}}_{6+\ensuremath{\delta}}$},\
  }\href {https://doi.org/10.1103/PhysRevB.89.184516} {\bibfield  {journal}
  {\bibinfo  {journal} {Phys. Rev. B}\ }\textbf {\bibinfo {volume} {89}},\
  \bibinfo {pages} {184516} (\bibinfo {year} {2014})}\BibitemShut {NoStop}%
\bibitem [{\citenamefont {Mitrano}\ \emph {et~al.}(2016)\citenamefont
  {Mitrano}, \citenamefont {Cantaluppi}, \citenamefont {Nicoletti},
  \citenamefont {Kaiser}, \citenamefont {Perucchi}, \citenamefont {Lupi},
  \citenamefont {Di~Pietro}, \citenamefont {Pontiroli}, \citenamefont {Ricco},
  \citenamefont {Clark}, \citenamefont {Jaksch},\ and\ \citenamefont
  {A.}}]{Nature.530.2016}%
  \BibitemOpen
  \bibfield  {author} {\bibinfo {author} {\bibfnamefont {M.}~\bibnamefont
  {Mitrano}}, \bibinfo {author} {\bibfnamefont {A.}~\bibnamefont {Cantaluppi}},
  \bibinfo {author} {\bibfnamefont {D.}~\bibnamefont {Nicoletti}}, \bibinfo
  {author} {\bibfnamefont {S.}~\bibnamefont {Kaiser}}, \bibinfo {author}
  {\bibfnamefont {A.}~\bibnamefont {Perucchi}}, \bibinfo {author}
  {\bibfnamefont {S.}~\bibnamefont {Lupi}}, \bibinfo {author} {\bibfnamefont
  {P.}~\bibnamefont {Di~Pietro}}, \bibinfo {author} {\bibfnamefont
  {D.}~\bibnamefont {Pontiroli}}, \bibinfo {author} {\bibfnamefont
  {M.}~\bibnamefont {Ricco}}, \bibinfo {author} {\bibfnamefont
  {S.}~\bibnamefont {Clark}}, \bibinfo {author} {\bibfnamefont
  {D.}~\bibnamefont {Jaksch}},\ and\ \bibinfo {author} {\bibfnamefont
  {C.}~\bibnamefont {A.}},\ }\bibfield  {title} {\bibinfo {title} {Possible
  light-induced superconductivity in k3c60 at high temperature},\ }\href
  {https://doi.org/10.1038/nature16522} {\bibfield  {journal} {\bibinfo
  {journal} {Nature}\ }\textbf {\bibinfo {volume} {530}},\ \bibinfo {pages}
  {461} (\bibinfo {year} {2016})}\BibitemShut {NoStop}%
\bibitem [{\citenamefont {Matsunaga}\ \emph {et~al.}(2013)\citenamefont
  {Matsunaga}, \citenamefont {Hamada}, \citenamefont {Makise}, \citenamefont
  {Uzawa}, \citenamefont {Terai}, \citenamefont {Wang},\ and\ \citenamefont
  {Shimano}}]{PhysRevLett.111.057002}%
  \BibitemOpen
  \bibfield  {author} {\bibinfo {author} {\bibfnamefont {R.}~\bibnamefont
  {Matsunaga}}, \bibinfo {author} {\bibfnamefont {Y.~I.}\ \bibnamefont
  {Hamada}}, \bibinfo {author} {\bibfnamefont {K.}~\bibnamefont {Makise}},
  \bibinfo {author} {\bibfnamefont {Y.}~\bibnamefont {Uzawa}}, \bibinfo
  {author} {\bibfnamefont {H.}~\bibnamefont {Terai}}, \bibinfo {author}
  {\bibfnamefont {Z.}~\bibnamefont {Wang}},\ and\ \bibinfo {author}
  {\bibfnamefont {R.}~\bibnamefont {Shimano}},\ }\bibfield  {title} {\bibinfo
  {title} {Higgs amplitude mode in the bcs superconductors
  ${\mathrm{nb}}_{1\mathrm{\text{\ensuremath{-}}}x}{\mathrm{ti}}_{x}\mathbf{N}$
  induced by terahertz pulse excitation},\ }\href
  {https://doi.org/10.1103/PhysRevLett.111.057002} {\bibfield  {journal}
  {\bibinfo  {journal} {Phys. Rev. Lett.}\ }\textbf {\bibinfo {volume} {111}},\
  \bibinfo {pages} {057002} (\bibinfo {year} {2013})}\BibitemShut {NoStop}%
\bibitem [{\citenamefont {Matsunaga}\ \emph {et~al.}(2014)\citenamefont
  {Matsunaga}, \citenamefont {Tsuji}, \citenamefont {Fujita}, \citenamefont
  {Sugioka}, \citenamefont {Makise}, \citenamefont {Uzawa}, \citenamefont
  {Terai}, \citenamefont {Wang}, \citenamefont {Aoki},\ and\ \citenamefont
  {Shimano}}]{Science.345.2014}%
  \BibitemOpen
  \bibfield  {author} {\bibinfo {author} {\bibfnamefont {R.}~\bibnamefont
  {Matsunaga}}, \bibinfo {author} {\bibfnamefont {N.}~\bibnamefont {Tsuji}},
  \bibinfo {author} {\bibfnamefont {H.}~\bibnamefont {Fujita}}, \bibinfo
  {author} {\bibfnamefont {A.}~\bibnamefont {Sugioka}}, \bibinfo {author}
  {\bibfnamefont {K.}~\bibnamefont {Makise}}, \bibinfo {author} {\bibfnamefont
  {Y.}~\bibnamefont {Uzawa}}, \bibinfo {author} {\bibfnamefont
  {H.}~\bibnamefont {Terai}}, \bibinfo {author} {\bibfnamefont
  {Z.}~\bibnamefont {Wang}}, \bibinfo {author} {\bibfnamefont {H.}~\bibnamefont
  {Aoki}},\ and\ \bibinfo {author} {\bibfnamefont {R.}~\bibnamefont
  {Shimano}},\ }\bibfield  {title} {\bibinfo {title} {Light-induced collective
  pseudospin precession resonating with higgs mode in a superconductor},\
  }\href {https://doi.org/10.1126/science.1254697} {\bibfield  {journal}
  {\bibinfo  {journal} {Science}\ }\textbf {\bibinfo {volume} {345}},\ \bibinfo
  {pages} {1145} (\bibinfo {year} {2014})},\ \Eprint
  {https://arxiv.org/abs/https://www.science.org/doi/pdf/10.1126/science.1254697}
  {https://www.science.org/doi/pdf/10.1126/science.1254697} \BibitemShut
  {NoStop}%
\bibitem [{\citenamefont {Matsunaga}\ \emph {et~al.}(2017)\citenamefont
  {Matsunaga}, \citenamefont {Tsuji}, \citenamefont {Makise}, \citenamefont
  {Terai}, \citenamefont {Aoki},\ and\ \citenamefont
  {Shimano}}]{PhysRevB.96.020505}%
  \BibitemOpen
  \bibfield  {author} {\bibinfo {author} {\bibfnamefont {R.}~\bibnamefont
  {Matsunaga}}, \bibinfo {author} {\bibfnamefont {N.}~\bibnamefont {Tsuji}},
  \bibinfo {author} {\bibfnamefont {K.}~\bibnamefont {Makise}}, \bibinfo
  {author} {\bibfnamefont {H.}~\bibnamefont {Terai}}, \bibinfo {author}
  {\bibfnamefont {H.}~\bibnamefont {Aoki}},\ and\ \bibinfo {author}
  {\bibfnamefont {R.}~\bibnamefont {Shimano}},\ }\bibfield  {title} {\bibinfo
  {title} {Polarization-resolved terahertz third-harmonic generation in a
  single-crystal superconductor nbn: Dominance of the higgs mode beyond the bcs
  approximation},\ }\href {https://doi.org/10.1103/PhysRevB.96.020505}
  {\bibfield  {journal} {\bibinfo  {journal} {Phys. Rev. B}\ }\textbf {\bibinfo
  {volume} {96}},\ \bibinfo {pages} {020505} (\bibinfo {year}
  {2017})}\BibitemShut {NoStop}%
\bibitem [{\citenamefont {Rohwer}\ \emph {et~al.}(2011)\citenamefont {Rohwer},
  \citenamefont {Hellmann}, \citenamefont {Wiesenmayer}, \citenamefont {Sohrt},
  \citenamefont {Stange}, \citenamefont {Slomski}, \citenamefont {Carr},
  \citenamefont {Liu}, \citenamefont {Avila}, \citenamefont {Kallane},
  \citenamefont {Mathias}, \citenamefont {Kipp}, \citenamefont {Rossnagel},\
  and\ \citenamefont {Bauer}}]{Nature.471.2011}%
  \BibitemOpen
  \bibfield  {author} {\bibinfo {author} {\bibfnamefont {T.}~\bibnamefont
  {Rohwer}}, \bibinfo {author} {\bibfnamefont {S.}~\bibnamefont {Hellmann}},
  \bibinfo {author} {\bibfnamefont {M.}~\bibnamefont {Wiesenmayer}}, \bibinfo
  {author} {\bibfnamefont {C.}~\bibnamefont {Sohrt}}, \bibinfo {author}
  {\bibfnamefont {A.}~\bibnamefont {Stange}}, \bibinfo {author} {\bibfnamefont
  {B.}~\bibnamefont {Slomski}}, \bibinfo {author} {\bibfnamefont
  {A.}~\bibnamefont {Carr}}, \bibinfo {author} {\bibfnamefont {Y.}~\bibnamefont
  {Liu}}, \bibinfo {author} {\bibfnamefont {L.~M.}\ \bibnamefont {Avila}},
  \bibinfo {author} {\bibfnamefont {M.}~\bibnamefont {Kallane}}, \bibinfo
  {author} {\bibfnamefont {S.}~\bibnamefont {Mathias}}, \bibinfo {author}
  {\bibfnamefont {L.}~\bibnamefont {Kipp}}, \bibinfo {author} {\bibfnamefont
  {K.}~\bibnamefont {Rossnagel}},\ and\ \bibinfo {author} {\bibfnamefont
  {M.}~\bibnamefont {Bauer}},\ }\bibfield  {title} {\bibinfo {title} {Collapse
  of long-range charge order tracked by time-resolved photoemission at high
  momenta},\ }\href {https://doi.org/10.1038/nature09829} {\bibfield  {journal}
  {\bibinfo  {journal} {Nature}\ }\textbf {\bibinfo {volume} {471}},\ \bibinfo
  {pages} {490} (\bibinfo {year} {2011})}\BibitemShut {NoStop}%
\bibitem [{\citenamefont {Hellmann}\ \emph {et~al.}(2012)\citenamefont
  {Hellmann}, \citenamefont {Rohwer}, \citenamefont {Kallane}, \citenamefont
  {Hanff}, \citenamefont {Sohrt}, \citenamefont {Stange}, \citenamefont {Carr},
  \citenamefont {Murnane}, \citenamefont {Kapteyn}, \citenamefont {Kipp},
  \citenamefont {Bauer},\ and\ \citenamefont {Rossnagel}}]{NatCommun.3.1069}%
  \BibitemOpen
  \bibfield  {author} {\bibinfo {author} {\bibfnamefont {S.}~\bibnamefont
  {Hellmann}}, \bibinfo {author} {\bibfnamefont {T.}~\bibnamefont {Rohwer}},
  \bibinfo {author} {\bibfnamefont {M.}~\bibnamefont {Kallane}}, \bibinfo
  {author} {\bibfnamefont {K.}~\bibnamefont {Hanff}}, \bibinfo {author}
  {\bibfnamefont {C.}~\bibnamefont {Sohrt}}, \bibinfo {author} {\bibfnamefont
  {A.}~\bibnamefont {Stange}}, \bibinfo {author} {\bibfnamefont
  {A.}~\bibnamefont {Carr}}, \bibinfo {author} {\bibfnamefont {M.}~\bibnamefont
  {Murnane}}, \bibinfo {author} {\bibfnamefont {H.}~\bibnamefont {Kapteyn}},
  \bibinfo {author} {\bibfnamefont {L.}~\bibnamefont {Kipp}}, \bibinfo {author}
  {\bibfnamefont {M.}~\bibnamefont {Bauer}},\ and\ \bibinfo {author}
  {\bibfnamefont {K.}~\bibnamefont {Rossnagel}},\ }\bibfield  {title} {\bibinfo
  {title} {Time-domain classification of charge-density-wave insulators},\
  }\href {https://doi.org/10.1038/ncomms2078} {\bibfield  {journal} {\bibinfo
  {journal} {Nat Commun}\ }\textbf {\bibinfo {volume} {3}},\ \bibinfo {pages}
  {1069} (\bibinfo {year} {2012})}\BibitemShut {NoStop}%
\bibitem [{\citenamefont {Porer}\ \emph {et~al.}(2014)\citenamefont {Porer},
  \citenamefont {Leierseder}, \citenamefont {Menard}, \citenamefont
  {Dachraoui}, \citenamefont {Mouchliadis}, \citenamefont {Perakis},
  \citenamefont {Heinzmann}, \citenamefont {Demsar}, \citenamefont {K.},\ and\
  \citenamefont {Huber}}]{NatureMater.13.857}%
  \BibitemOpen
  \bibfield  {author} {\bibinfo {author} {\bibfnamefont {M.}~\bibnamefont
  {Porer}}, \bibinfo {author} {\bibfnamefont {U.}~\bibnamefont {Leierseder}},
  \bibinfo {author} {\bibfnamefont {J.-M.}\ \bibnamefont {Menard}}, \bibinfo
  {author} {\bibfnamefont {H.}~\bibnamefont {Dachraoui}}, \bibinfo {author}
  {\bibfnamefont {L.}~\bibnamefont {Mouchliadis}}, \bibinfo {author}
  {\bibfnamefont {I.~E.}\ \bibnamefont {Perakis}}, \bibinfo {author}
  {\bibfnamefont {U.}~\bibnamefont {Heinzmann}}, \bibinfo {author}
  {\bibfnamefont {J.}~\bibnamefont {Demsar}}, \bibinfo {author} {\bibfnamefont
  {R.}~\bibnamefont {K.}},\ and\ \bibinfo {author} {\bibfnamefont
  {R.}~\bibnamefont {Huber}},\ }\bibfield  {title} {\bibinfo {title}
  {Non-thermal separation of electronic and structural orders in a persisting
  charge density wave},\ }\href {https://doi.org/10.1038/nmat4042} {\bibfield
  {journal} {\bibinfo  {journal} {Nature Mater}\ }\textbf {\bibinfo {volume}
  {13}},\ \bibinfo {pages} {857} (\bibinfo {year} {2014})}\BibitemShut
  {NoStop}%
\bibitem [{\citenamefont {Gole\ifmmode~\check{z}\else \v{z}\fi{}}\ \emph
  {et~al.}(2016)\citenamefont {Gole\ifmmode~\check{z}\else \v{z}\fi{}},
  \citenamefont {Werner},\ and\ \citenamefont {Eckstein}}]{PhysRevB.94.035121}%
  \BibitemOpen
  \bibfield  {author} {\bibinfo {author} {\bibfnamefont {D.}~\bibnamefont
  {Gole\ifmmode~\check{z}\else \v{z}\fi{}}}, \bibinfo {author} {\bibfnamefont
  {P.}~\bibnamefont {Werner}},\ and\ \bibinfo {author} {\bibfnamefont
  {M.}~\bibnamefont {Eckstein}},\ }\bibfield  {title} {\bibinfo {title}
  {Photoinduced gap closure in an excitonic insulator},\ }\href
  {https://doi.org/10.1103/PhysRevB.94.035121} {\bibfield  {journal} {\bibinfo
  {journal} {Phys. Rev. B}\ }\textbf {\bibinfo {volume} {94}},\ \bibinfo
  {pages} {035121} (\bibinfo {year} {2016})}\BibitemShut {NoStop}%
\bibitem [{\citenamefont {Mor}\ \emph {et~al.}(2017)\citenamefont {Mor},
  \citenamefont {Herzog}, \citenamefont {Gole\ifmmode~\check{z}\else
  \v{z}\fi{}}, \citenamefont {Werner}, \citenamefont {Eckstein}, \citenamefont
  {Katayama}, \citenamefont {Nohara}, \citenamefont {Takagi}, \citenamefont
  {Mizokawa}, \citenamefont {Monney},\ and\ \citenamefont
  {St\"ahler}}]{PhysRevLett.119.086401}%
  \BibitemOpen
  \bibfield  {author} {\bibinfo {author} {\bibfnamefont {S.}~\bibnamefont
  {Mor}}, \bibinfo {author} {\bibfnamefont {M.}~\bibnamefont {Herzog}},
  \bibinfo {author} {\bibfnamefont {D.}~\bibnamefont
  {Gole\ifmmode~\check{z}\else \v{z}\fi{}}}, \bibinfo {author} {\bibfnamefont
  {P.}~\bibnamefont {Werner}}, \bibinfo {author} {\bibfnamefont
  {M.}~\bibnamefont {Eckstein}}, \bibinfo {author} {\bibfnamefont
  {N.}~\bibnamefont {Katayama}}, \bibinfo {author} {\bibfnamefont
  {M.}~\bibnamefont {Nohara}}, \bibinfo {author} {\bibfnamefont
  {H.}~\bibnamefont {Takagi}}, \bibinfo {author} {\bibfnamefont
  {T.}~\bibnamefont {Mizokawa}}, \bibinfo {author} {\bibfnamefont
  {C.}~\bibnamefont {Monney}},\ and\ \bibinfo {author} {\bibfnamefont
  {J.}~\bibnamefont {St\"ahler}},\ }\bibfield  {title} {\bibinfo {title}
  {Ultrafast electronic band gap control in an excitonic insulator},\ }\href
  {https://doi.org/10.1103/PhysRevLett.119.086401} {\bibfield  {journal}
  {\bibinfo  {journal} {Phys. Rev. Lett.}\ }\textbf {\bibinfo {volume} {119}},\
  \bibinfo {pages} {086401} (\bibinfo {year} {2017})}\BibitemShut {NoStop}%
\bibitem [{\citenamefont {Werdehausen}\ \emph {et~al.}(2018)\citenamefont
  {Werdehausen}, \citenamefont {Takayama}, \citenamefont {Hoppner},
  \citenamefont {Albrecht}, \citenamefont {Rost}, \citenamefont {Lu},
  \citenamefont {Manske}, \citenamefont {Takagi},\ and\ \citenamefont
  {Kaiser}}]{Werdehausen_2018}%
  \BibitemOpen
  \bibfield  {author} {\bibinfo {author} {\bibfnamefont {D.}~\bibnamefont
  {Werdehausen}}, \bibinfo {author} {\bibfnamefont {T.}~\bibnamefont
  {Takayama}}, \bibinfo {author} {\bibfnamefont {M.}~\bibnamefont {Hoppner}},
  \bibinfo {author} {\bibfnamefont {G.}~\bibnamefont {Albrecht}}, \bibinfo
  {author} {\bibfnamefont {A.~W.}\ \bibnamefont {Rost}}, \bibinfo {author}
  {\bibfnamefont {Y.}~\bibnamefont {Lu}}, \bibinfo {author} {\bibfnamefont
  {D.}~\bibnamefont {Manske}}, \bibinfo {author} {\bibfnamefont
  {H.}~\bibnamefont {Takagi}},\ and\ \bibinfo {author} {\bibfnamefont
  {S.}~\bibnamefont {Kaiser}},\ }\bibfield  {title} {\bibinfo {title} {Coherent
  order parameter oscillations in the ground state of the excitonic insulator
  ta$_2$nise$_5$},\ }\bibfield  {journal} {\bibinfo  {journal} {Science
  Advances}\ }\textbf {\bibinfo {volume} {4}},\ \href
  {https://doi.org/10.1126/sciadv.aap8652} {10.1126/sciadv.aap8652} (\bibinfo
  {year} {2018})\BibitemShut {NoStop}%
\bibitem [{\citenamefont {Kune{\v{s}}}(2015)}]{JPhysCondensMatter.27.333201}%
  \BibitemOpen
  \bibfield  {author} {\bibinfo {author} {\bibfnamefont {J.}~\bibnamefont
  {Kune{\v{s}}}},\ }\bibfield  {title} {\bibinfo {title} {Excitonic
  condensation in systems of strongly correlated electrons},\ }\href
  {https://doi.org/10.1088/0953-8984/27/33/333201} {\bibfield  {journal}
  {\bibinfo  {journal} {J. Phys. Condens. Matter}\ }\textbf {\bibinfo {volume}
  {27}},\ \bibinfo {pages} {333201} (\bibinfo {year} {2015})}\BibitemShut
  {NoStop}%
\bibitem [{\citenamefont {Werner}\ and\ \citenamefont
  {Millis}(2007)}]{PhysRevLett.99.126405}%
  \BibitemOpen
  \bibfield  {author} {\bibinfo {author} {\bibfnamefont {P.}~\bibnamefont
  {Werner}}\ and\ \bibinfo {author} {\bibfnamefont {A.~J.}\ \bibnamefont
  {Millis}},\ }\bibfield  {title} {\bibinfo {title} {High-spin to low-spin and
  orbital polarization transitions in multiorbital mott systems},\ }\href
  {https://doi.org/10.1103/PhysRevLett.99.126405} {\bibfield  {journal}
  {\bibinfo  {journal} {Phys. Rev. Lett.}\ }\textbf {\bibinfo {volume} {99}},\
  \bibinfo {pages} {126405} (\bibinfo {year} {2007})}\BibitemShut {NoStop}%
\bibitem [{\citenamefont {Suzuki}\ \emph {et~al.}(2009)\citenamefont {Suzuki},
  \citenamefont {Watanabe},\ and\ \citenamefont
  {Ishihara}}]{PhysRevB.80.054410}%
  \BibitemOpen
  \bibfield  {author} {\bibinfo {author} {\bibfnamefont {R.}~\bibnamefont
  {Suzuki}}, \bibinfo {author} {\bibfnamefont {T.}~\bibnamefont {Watanabe}},\
  and\ \bibinfo {author} {\bibfnamefont {S.}~\bibnamefont {Ishihara}},\
  }\bibfield  {title} {\bibinfo {title} {Spin-state transition and phase
  separation in a multiorbital hubbard model},\ }\href
  {https://doi.org/10.1103/PhysRevB.80.054410} {\bibfield  {journal} {\bibinfo
  {journal} {Phys. Rev. B}\ }\textbf {\bibinfo {volume} {80}},\ \bibinfo
  {pages} {054410} (\bibinfo {year} {2009})}\BibitemShut {NoStop}%
\bibitem [{\citenamefont {Balents}(2000)}]{PhysRevB.62.2346}%
  \BibitemOpen
  \bibfield  {author} {\bibinfo {author} {\bibfnamefont {L.}~\bibnamefont
  {Balents}},\ }\bibfield  {title} {\bibinfo {title} {Excitonic order at strong
  coupling: Pseudospin, doping, and ferromagnetism},\ }\href
  {https://doi.org/10.1103/PhysRevB.62.2346} {\bibfield  {journal} {\bibinfo
  {journal} {Phys. Rev. B}\ }\textbf {\bibinfo {volume} {62}},\ \bibinfo
  {pages} {2346} (\bibinfo {year} {2000})}\BibitemShut {NoStop}%
\bibitem [{\citenamefont {Kaneko}\ and\ \citenamefont
  {Ohta}(2014)}]{PhysRevB.90.245144}%
  \BibitemOpen
  \bibfield  {author} {\bibinfo {author} {\bibfnamefont {T.}~\bibnamefont
  {Kaneko}}\ and\ \bibinfo {author} {\bibfnamefont {Y.}~\bibnamefont {Ohta}},\
  }\bibfield  {title} {\bibinfo {title} {Roles of hund's rule coupling in
  excitonic density-wave states},\ }\href
  {https://doi.org/10.1103/PhysRevB.90.245144} {\bibfield  {journal} {\bibinfo
  {journal} {Phys. Rev. B}\ }\textbf {\bibinfo {volume} {90}},\ \bibinfo
  {pages} {245144} (\bibinfo {year} {2014})}\BibitemShut {NoStop}%
\bibitem [{\citenamefont {Kune{\v{s}}}\ and\ \citenamefont
  {Augustinsk\'y}(2014)}]{PhysRevB.89.115134}%
  \BibitemOpen
  \bibfield  {author} {\bibinfo {author} {\bibfnamefont {J.}~\bibnamefont
  {Kune{\v{s}}}}\ and\ \bibinfo {author} {\bibfnamefont {P.}~\bibnamefont
  {Augustinsk\'y}},\ }\bibfield  {title} {\bibinfo {title} {Excitonic
  instability at the spin-state transition in the two-band hubbard model},\
  }\href {https://doi.org/10.1103/PhysRevB.89.115134} {\bibfield  {journal}
  {\bibinfo  {journal} {Phys. Rev. B}\ }\textbf {\bibinfo {volume} {89}},\
  \bibinfo {pages} {115134} (\bibinfo {year} {2014})}\BibitemShut {NoStop}%
\bibitem [{\citenamefont {Murakami}\ \emph {et~al.}(2017)\citenamefont
  {Murakami}, \citenamefont {Gole\ifmmode~\check{z}\else \v{z}\fi{}},
  \citenamefont {Eckstein},\ and\ \citenamefont
  {Werner}}]{PhysRevLett.119.247601}%
  \BibitemOpen
  \bibfield  {author} {\bibinfo {author} {\bibfnamefont {Y.}~\bibnamefont
  {Murakami}}, \bibinfo {author} {\bibfnamefont {D.}~\bibnamefont
  {Gole\ifmmode~\check{z}\else \v{z}\fi{}}}, \bibinfo {author} {\bibfnamefont
  {M.}~\bibnamefont {Eckstein}},\ and\ \bibinfo {author} {\bibfnamefont
  {P.}~\bibnamefont {Werner}},\ }\bibfield  {title} {\bibinfo {title}
  {Photoinduced enhancement of excitonic order},\ }\href
  {https://doi.org/10.1103/PhysRevLett.119.247601} {\bibfield  {journal}
  {\bibinfo  {journal} {Phys. Rev. Lett.}\ }\textbf {\bibinfo {volume} {119}},\
  \bibinfo {pages} {247601} (\bibinfo {year} {2017})}\BibitemShut {NoStop}%
\bibitem [{\citenamefont {Kanamori}(1963)}]{Kanamori1963}%
  \BibitemOpen
  \bibfield  {author} {\bibinfo {author} {\bibfnamefont {J.}~\bibnamefont
  {Kanamori}},\ }\bibfield  {title} {\bibinfo {title} {Electron correlation and
  ferromagnetism of transition metals},\ }\href@noop {} {\bibfield  {journal}
  {\bibinfo  {journal} {Prog. Theor. Phys.}\ }\textbf {\bibinfo {volume}
  {30}},\ \bibinfo {pages} {275} (\bibinfo {year} {1963})}\BibitemShut
  {NoStop}%
\bibitem [{\citenamefont {Orlov}\ \emph {et~al.}(2021)\citenamefont {Orlov},
  \citenamefont {Nikolaev}, \citenamefont {Dudnikov},\ and\ \citenamefont
  {Ovchinnikov}}]{Orlov_PhysRevB.104.195103}%
  \BibitemOpen
  \bibfield  {author} {\bibinfo {author} {\bibfnamefont {Y.~S.}\ \bibnamefont
  {Orlov}}, \bibinfo {author} {\bibfnamefont {S.~V.}\ \bibnamefont {Nikolaev}},
  \bibinfo {author} {\bibfnamefont {V.~A.}\ \bibnamefont {Dudnikov}},\ and\
  \bibinfo {author} {\bibfnamefont {S.~G.}\ \bibnamefont {Ovchinnikov}},\
  }\bibfield  {title} {\bibinfo {title} {Forming a dielectric exciton phase in
  strongly correlated systems with spin crossover},\ }\href
  {https://doi.org/10.1103/PhysRevB.104.195103} {\bibfield  {journal} {\bibinfo
   {journal} {Phys. Rev. B}\ }\textbf {\bibinfo {volume} {104}},\ \bibinfo
  {pages} {195103} (\bibinfo {year} {2021})}\BibitemShut {NoStop}%
\bibitem [{\citenamefont {Hubbard}(1964)}]{Hubbard1964}%
  \BibitemOpen
  \bibfield  {author} {\bibinfo {author} {\bibfnamefont {J.}~\bibnamefont
  {Hubbard}},\ }\bibfield  {title} {\bibinfo {title} {Electron correlations in
  narrow energy bands. ii. the degenerate band case},\ }\href
  {https://doi.org/10.1098/rspa.1964.0019} {\bibfield  {journal} {\bibinfo
  {journal} {Proc. R. Soc. A}\ }\textbf {\bibinfo {volume} {277}},\ \bibinfo
  {pages} {237} (\bibinfo {year} {1964})}\BibitemShut {NoStop}%
\bibitem [{\citenamefont {Zaitsev}(1976)}]{Zaitsev1976}%
  \BibitemOpen
  \bibfield  {author} {\bibinfo {author} {\bibfnamefont {R.~O.}\ \bibnamefont
  {Zaitsev}},\ }\bibfield  {title} {\bibinfo {title} {Diagram technique and gas
  approximation in the hubbard model},\ }\href
  {http://www.jetp.ac.ru/cgi-bin/e/index/e/43/3/p574?a=list} {\bibfield
  {journal} {\bibinfo  {journal} {Sov. Phys. JETP}\ }\textbf {\bibinfo {volume}
  {43}},\ \bibinfo {pages} {574} (\bibinfo {year} {1976})}\BibitemShut
  {NoStop}%
\bibitem [{\citenamefont {Chao}\ \emph {et~al.}(1977)\citenamefont {Chao},
  \citenamefont {Spalek},\ and\ \citenamefont {Oles}}]{Chao1977}%
  \BibitemOpen
  \bibfield  {author} {\bibinfo {author} {\bibfnamefont {K.~A.}\ \bibnamefont
  {Chao}}, \bibinfo {author} {\bibfnamefont {J.}~\bibnamefont {Spalek}},\ and\
  \bibinfo {author} {\bibfnamefont {A.~M.}\ \bibnamefont {Oles}},\ }\bibfield
  {title} {\bibinfo {title} {Kinetic exchange interaction in a narrow s-band},\
  }\href {https://doi.org/10.1088/0022-3719/10/10/002} {\bibfield  {journal}
  {\bibinfo  {journal} {J. Phys. C}\ }\textbf {\bibinfo {volume} {10}},\
  \bibinfo {pages} {L271} (\bibinfo {year} {1977})}\BibitemShut {NoStop}%
\bibitem [{\citenamefont {Gavrichkov}\ \emph {et~al.}(2017)\citenamefont
  {Gavrichkov}, \citenamefont {Polukeev},\ and\ \citenamefont
  {Ovchinnikov}}]{Gavrichkov_Polukeev}%
  \BibitemOpen
  \bibfield  {author} {\bibinfo {author} {\bibfnamefont {V.~A.}\ \bibnamefont
  {Gavrichkov}}, \bibinfo {author} {\bibfnamefont {S.~I.}\ \bibnamefont
  {Polukeev}},\ and\ \bibinfo {author} {\bibfnamefont {S.~G.}\ \bibnamefont
  {Ovchinnikov}},\ }\bibfield  {title} {\bibinfo {title} {Contribution from
  optically excited many-electron states to the superexchange interaction in
  mott-hubbard insulators},\ }\href
  {https://doi.org/10.1103/PhysRevB.95.144424} {\bibfield  {journal} {\bibinfo
  {journal} {Phys. Rev. B}\ }\textbf {\bibinfo {volume} {95}},\ \bibinfo
  {pages} {144424} (\bibinfo {year} {2017})}\BibitemShut {NoStop}%
\bibitem [{\citenamefont {Val'kov}\ and\ \citenamefont
  {Ovchinnikov}(1982)}]{Valkov}%
  \BibitemOpen
  \bibfield  {author} {\bibinfo {author} {\bibfnamefont {V.~V.}\ \bibnamefont
  {Val'kov}}\ and\ \bibinfo {author} {\bibfnamefont {S.~G.}\ \bibnamefont
  {Ovchinnikov}},\ }\bibfield  {title} {\bibinfo {title} {Hubbard operators and
  spin-wave theory of heisenberg magnets with arbitrary spin},\ }\href
  {https://doi.org/10.1007/BF01016463} {\bibfield  {journal} {\bibinfo
  {journal} {Theor. Math. Phys.}\ }\textbf {\bibinfo {volume} {50}},\ \bibinfo
  {pages} {466} (\bibinfo {year} {1982})}\BibitemShut {NoStop}%
\bibitem [{\citenamefont {Vonsovskii}\ and\ \citenamefont
  {Svirskii}(1965)}]{Vonsovsky}%
  \BibitemOpen
  \bibfield  {author} {\bibinfo {author} {\bibfnamefont {S.~V.}\ \bibnamefont
  {Vonsovskii}}\ and\ \bibinfo {author} {\bibfnamefont {M.~S.}\ \bibnamefont
  {Svirskii}},\ }\bibfield  {title} {\bibinfo {title} {The effect of the
  multiplicity of d (f) shells on electron interaction in crystals},\
  }\href@noop {} {\bibfield  {journal} {\bibinfo  {journal} {Sov. Phys. JETP}\
  }\textbf {\bibinfo {volume} {20}} (\bibinfo {year} {1965})}\BibitemShut
  {NoStop}%
\bibitem [{\citenamefont {Nasu}\ and\ \citenamefont
  {Ishihara}(2013)}]{PhysRevB.88.205110}%
  \BibitemOpen
  \bibfield  {author} {\bibinfo {author} {\bibfnamefont {J.}~\bibnamefont
  {Nasu}}\ and\ \bibinfo {author} {\bibfnamefont {S.}~\bibnamefont
  {Ishihara}},\ }\bibfield  {title} {\bibinfo {title} {Vibronic excitation
  dynamics in orbitally degenerate correlated electron system},\ }\href
  {https://doi.org/10.1103/PhysRevB.88.205110} {\bibfield  {journal} {\bibinfo
  {journal} {Phys. Rev. B}\ }\textbf {\bibinfo {volume} {88}},\ \bibinfo
  {pages} {205110} (\bibinfo {year} {2013})}\BibitemShut {NoStop}%
\bibitem [{\citenamefont {Onufrieva}(1985)}]{Onufrieva}%
  \BibitemOpen
  \bibfield  {author} {\bibinfo {author} {\bibfnamefont {F.~P.}\ \bibnamefont
  {Onufrieva}},\ }\bibfield  {title} {\bibinfo {title} {Low-temperature
  properties of spin systems with tensor order parameter},\ }\href@noop {}
  {\bibfield  {journal} {\bibinfo  {journal} {Sov. Phys. JETP}\ }\textbf
  {\bibinfo {volume} {62}},\ \bibinfo {pages} {1311} (\bibinfo {year}
  {1985})}\BibitemShut {NoStop}%
\bibitem [{\citenamefont {Colpa}(1978)}]{Colpa78}%
  \BibitemOpen
  \bibfield  {author} {\bibinfo {author} {\bibfnamefont {J.}~\bibnamefont
  {Colpa}},\ }\bibfield  {title} {\bibinfo {title} {Diagonalisation of the
  quadratic boson hamiltonian},\ }\href@noop {} {\bibfield  {journal} {\bibinfo
   {journal} {Physica}\ }\textbf {\bibinfo {volume} {93A}},\ \bibinfo {pages}
  {327} (\bibinfo {year} {1978})}\BibitemShut {NoStop}%
\bibitem [{\citenamefont {Hoch}\ \emph {et~al.}(2009)\citenamefont {Hoch},
  \citenamefont {Nellutla}, \citenamefont {van Tol}, \citenamefont {Choi},
  \citenamefont {Lu}, \citenamefont {Zheng},\ and\ \citenamefont
  {Mitchell}}]{PhysRevB.79.214421}%
  \BibitemOpen
  \bibfield  {author} {\bibinfo {author} {\bibfnamefont {M.~J.~R.}\
  \bibnamefont {Hoch}}, \bibinfo {author} {\bibfnamefont {S.}~\bibnamefont
  {Nellutla}}, \bibinfo {author} {\bibfnamefont {J.}~\bibnamefont {van Tol}},
  \bibinfo {author} {\bibfnamefont {E.~S.}\ \bibnamefont {Choi}}, \bibinfo
  {author} {\bibfnamefont {J.}~\bibnamefont {Lu}}, \bibinfo {author}
  {\bibfnamefont {H.}~\bibnamefont {Zheng}},\ and\ \bibinfo {author}
  {\bibfnamefont {J.~F.}\ \bibnamefont {Mitchell}},\ }\bibfield  {title}
  {\bibinfo {title} {Diamagnetic to paramagnetic transition in
  ${\text{lacoo}}_{3}$},\ }\href {https://doi.org/10.1103/PhysRevB.79.214421}
  {\bibfield  {journal} {\bibinfo  {journal} {Phys. Rev. B}\ }\textbf {\bibinfo
  {volume} {79}},\ \bibinfo {pages} {214421} (\bibinfo {year}
  {2009})}\BibitemShut {NoStop}%
\bibitem [{\citenamefont {Blum}(2012)}]{Blum}%
  \BibitemOpen
  \bibfield  {author} {\bibinfo {author} {\bibfnamefont {K.}~\bibnamefont
  {Blum}},\ }\href@noop {} {\emph {\bibinfo {title} {Density Matrix Theory and
  Applications}}}\ (\bibinfo  {publisher} {Springer-Verlag},\ \bibinfo
  {address} {Berlin, Heidelberg},\ \bibinfo {year} {2012})\BibitemShut
  {NoStop}%
\bibitem [{\citenamefont {Orlov}\ \emph {et~al.}(2022)\citenamefont {Orlov},
  \citenamefont {Nikolaev}, \citenamefont {Kuz'min}, \citenamefont {Zarubin},\
  and\ \citenamefont {Ovchinnikov}}]{Exciton_arxiv}%
  \BibitemOpen
  \bibfield  {author} {\bibinfo {author} {\bibfnamefont {Y.~S.}\ \bibnamefont
  {Orlov}}, \bibinfo {author} {\bibfnamefont {S.~V.}\ \bibnamefont {Nikolaev}},
  \bibinfo {author} {\bibfnamefont {V.~I.}\ \bibnamefont {Kuz'min}}, \bibinfo
  {author} {\bibfnamefont {A.~E.}\ \bibnamefont {Zarubin}},\ and\ \bibinfo
  {author} {\bibfnamefont {S.~G.}\ \bibnamefont {Ovchinnikov}},\ }\href
  {https://doi.org/10.48550/ARXIV.2206.13821} {\bibinfo {title} {Excitonic
  ordering in strongly correlated spin crossover systems: induced magnetism and
  excitonic excitation spectrum}} (\bibinfo {year} {2022})\BibitemShut
  {NoStop}%
\end{thebibliography}%

\end{document}